\documentclass[pdflatex,sn-apa]{sn-jnl}

\usepackage{graphicx}%
\usepackage{multirow}%
\usepackage{amsmath,amssymb,amsfonts}%
\usepackage{amsthm}%
 \usepackage{hhline}%
\usepackage{mathrsfs}%
\usepackage[title]{appendix}%
\usepackage{xcolor}%
\usepackage{float}
\usepackage{textcomp}%
\usepackage{manyfoot}%
\usepackage{booktabs}%
\usepackage{algorithm}%
\usepackage{algorithmicx}%
\usepackage{algpseudocode}%
\usepackage{listings}%

\theoremstyle{thmstyleone}%
\theoremstyle{thmstyletwo}%

\theoremstyle{thmstylethree}%

\begin{document}

\title[A new functional model for partially observed functional data]{A novel generalized additive scalar-on-function regression model for partially observed multidimensional functional data:  an application to air quality classification}


\author*[1]{\fnm{Pavel} \sur{Hernández-Amaro}}\email{pahernan@est-econ.uc3m.es}

\author[1]{\fnm{Maria} \sur{Durban}}\email{mdurban@est-econ.uc3m.es}
\equalcont{These authors contributed equally to this work.}

\author[2]{\fnm{M. Carmen} \sur{Aguilera-Morillo}}\email{mdagumor@eio.upv.es}
\equalcont{These authors contributed equally to this work.}

\affil*[1]{\orgdiv{Estadística y Econometría}, \orgname{Universidad Carlos III de Madrid}, \orgaddress{\street{Avda. Universidad, 31}, \city{Leganes}, \postcode{28911}, \state{Madrid}, \country{España}}}

\affil[2]{\orgdiv{Estadística e Investigación Operativa Aplicadas y Calidad}, \orgname{Universitat Politècnica de València}, \orgaddress{\street{Camino de Vera, s/n}, \city{Valencia}, \postcode{46022}, \state{Valencia}, \country{España}}}


\abstract{In this work we propose a generalized additive functional regression model for partially observed functional data. Our approach accommodates functional predictors of varying dimensions without requiring imputation of missing observations. Both the functional coefficients and covariates are represented using basis function expansions, with B-splines used in this study, though the method is not restricted to any specific basis choice. Model coefficients are estimated via penalized likelihood, leveraging the mixed model representation of penalized splines for efficient computation and smoothing parameter estimation.The performance of the proposed approach is assessed through two simulation studies: one involving two one-dimensional functional covariates, and another using a two-dimensional functional covariate. Finally, we demonstrate the practical utility of our method in an application to air‑pollution classification in Dimapur, India, where images are treated as observations of a two-dimensional functional variable. This case study highlights the model’s ability to effectively handle incomplete functional data and to accurately discriminate between pollution levels.}




\maketitle

\section{Introduction}\label{sec1}

In the context of functional regression models, it is typically assumed that the sample data are fully observed over their entire domain. However, this assumption is often unrealistic in practical applications. In many real-world scenarios, the observed functional data are incomplete or only partially available, containing gaps or missing segments across their domains. Such situations arise frequently, as documented in studies such as \cite{Delaigle2016ApproximatingChains, Zhenhua2021BasisSnippets}. This type of data, commonly referred to as partially observed functional data, has been discussed under various terminologies in the literature; a comprehensive review of these terms and their associated methodologies can be found in \cite{Matthews2021Shape-BasedAnthropology}.

A substantial portion of the existing literature on partially observed functional data focuses on reconstruction, that is, estimating the missing segments of the functions before proceeding with statistical analysis. Examples of such gap-filling approaches for one-dimensional functional data include \cite{Delaigle2013ClassificationData, Kraus2015ComponentsData, Kneip2020OnData}. These methods often involve estimating functional principal components (FPCs) from the observed data and using them to reconstruct the curves. However, these techniques have been largely restricted to one-dimensional functional data, where each observation is a function of a single variable (e.g., time). In contrast, when dealing with two-dimensional functional data, such as images or spatio-temporal domains, the literature is much sparser, and gap-filling methods typically do not adopt a functional data analysis (FDA) perspective. Instead, these approaches are rooted in time series analysis or image processing techniques, particularly in the context of remote sensing and satellite imagery. Representative examples include \cite{Chen2011AImages, Cai2017PerformanceData, Zhang2014ApplicationImagery}, where various statistical and machine learning-based methods have been proposed to impute missing pixels or segments in images. These methods, however, do not fully exploit the smoothness and functional nature of the data and, in most of them, the time-series structure of pixels is used, so each image is required at different times.

Moreover, there are very few references to functional regression models applied to this type of data, regardless of its dimensionality. In most cases, the common practice is to perform a preliminary step of gap-filling before fitting the model. To the best of our knowledge, the only reference that proposes a functional regression model for this type of data without gap-filling can be found in \cite{Wang2022FunctionalData}. In this work, the authors propose a method that first estimates the principal components of partially observed functional data and then uses them to fit a functional regression model. However, there are already well‑established methodologies for estimating principal components from partially observed functional data, as referenced in the paper itself (e.g. \cite{Kraus2015ComponentsData} and \cite{Yao2005FunctionalData}). In addition, the proposed approach is limited to a single functional covariate and does not address the more general setting of an additive model. Existing methods typically represent functional data as one-dimensional curves and therefore do not naturally generalize to higher-dimensional functional domains. Furthermore, the additive regression framework, which allows flexible and interpretable modeling of multiple functional covariates, has not been explored for partially observed, multidimensional functional data.

To address these gaps in the literature, we propose a novel generalized multidimensional additive scalar-on-function regression model designed specifically for partially observed functional data. Our methodology makes the following key contributions: 
\begin{enumerate}
\item No need for data imputation, i.e., the proposed model estimates the regression coefficients directly from the partially observed data eliminating the need for an additional imputation step.

\item  The approach retains the intrinsic smoothness of functional data, enabling natural 
denoising of discrete observations and preserving the continuous nature of the underlying 
functions. We extend this property to partially observed functional covariates through a 
penalized weighted least squares estimation of their basis coefficients.

\item The functional regression coefficients are estimated using B-spline basis functions with adaptable roughness penalties, allowing for flexible and interpretable function estimation.

\item The model accommodates multiple functional covariates of varying dimensions, including two-dimensional surfaces, making it applicable to a wide range of complex data types.
\end{enumerate}
To the best of our knowledge, this is the first work to propose an additive functional regression framework capable of handling partially observed data with functional covariates in more than one dimension.

The remainder of the paper is organized as follows: In section \ref{section model} , we formally define partially observed functional data and introduce our proposed regression model along with its estimation procedure. Section \ref{section sim} presents two simulation studies that evaluate the performance of our approach compared to a gap-filling technique extended to the two-dimensional setting, demonstrating the advantages of our method. Section \ref{section data} applies the proposed methodology to a real-world air quality classification problem. Finally, Section \ref{section conclusion} offers concluding remarks and outlines potential directions for future research.

\section{Functional regression model for Partially Observed Multidimensional Functional Data}\label{section model}

Given the following sample data: $\{Y_i, \boldsymbol{X}_i(t)\}$, $ i=1, \ldots, N$, where the response variable $Y_i$ is a scalar outcome following an exponential family distribution with mean $\mu_i,$ $\boldsymbol{X}_i(\boldsymbol{t})$ is a vector of functions $\boldsymbol{X}_i(\boldsymbol{t}) = (X_i^{(1)}(t_{1}), X_i^{(2)}(t_{2}), \ldots, X_i^{(k)}(t_{k}))$, and $\boldsymbol{t} = (t_1,t_2,\ldots, t_k)\in \mathbb{D} = \mathbb{D}^{1} \times \mathbb{D}^{2} \times \ldots \times \mathbb{D}^{k} $ is a $k$-tuple of $d_1,d_2, \ldots, d_k$ dimensional vectors, each function $X^{(j)}(t_j)$ is defined in a domain $\mathbb{D}^j$ with inner product:
\begin{equation*}
    \langle f^{(j)},g^{(j)} \rangle = \displaystyle\int\limits_{D^j}f^{(j)}(t_j)g^{(j)}(t_j)\text{d}t_j,
\end{equation*}
which induces a Hilbert space structure on the functional space.

In practice, the functional predictors $X_i^{(j)}(t_j)$ may not be fully observed on their domains $\mathbb{D}^j$, but only on subject-specific subdomains $\mathbb{D}_i^j \subseteq \mathbb{D}^j$. This incomplete observation may arise due to instrument limitations, study dropout, or design constraints. If the region where the data is  missing is small and not at the boundaries of the domain, interpolation techniques could be used to reconstruct the sample functions $X_i^{j}(t_j)$ over the complete domain. However, when the missing regions are located at the domain boundaries, these methods are unreliable. To address this, we propose the following additive scalar-on-function regression model for partially observed multidimensional functional data:
\begin{equation}
\label{eq:model}
    g(\mu_i) = \alpha + \displaystyle \sum\limits_{j=1}^k \int\limits_{D_i^j} X_i^{(j)}(t_j) \beta^{(j)}(t_j)\text{d}t_j, \quad t_j\in \mathbb{D}_i^j\subset \mathbb{R}^{d_j}, \quad i =1, \ldots, N,
\end{equation}
where $\alpha$ is the intercept and $\beta^{(j)}(t_j)$ is the unknown functional coefficient associated with the $j$-th functional covariate.

The key distinction between this model and the conventional scalar-on-function regression model is that each functional covariate is evaluated only over its observed domain $\mathbb{D}_i^j$. Only the observed portions of the functions contribute information to the model, as integration is performed solely over these domains for each functional covariate.  Moreover, functional covariates with different dimensions can be included in the proposed model.

As it is well known, in practice we  have discrete observations of the sample functions that usually are affected by measurements errors or noise. To address these issues, the next section describes the estimation of model (\ref{eq:model}) under the assumption that both the functional covariates and the functional coefficients can be represented through basis expansions.

\subsection{Model formulation in terms of basis functions}

\label{2.1}

To estimate the model, we represent both the functional predictors and their corresponding coefficient functions using finite basis expansions. This approach transforms the original functional regression problem into a finite‑dimensional representation while preserving the functional nature of the data. Specifically, for each functional predictor of dimension $d_j$, and its associated coefficient function, we express them as linear combinations of selected basis functions, as described below:
\begin{eqnarray*}
	X_i^{(j)}(t_j) &=& \displaystyle \sum_{l=1}^{\delta_j}a_{il}^{(j)}\Phi_{l}(t_j),\\
	 \beta^{(j)}(t_j)  &=&  \sum_{v=1}^{s_j}b^{(j)}_{_v}\Gamma_v(t_j),
\end{eqnarray*}

where $\boldsymbol{\Phi}_l(t_j)$ and $\boldsymbol{\Gamma}_v(t_j)$ are tensor products of 
$d_j$ univariate B-spline basis functions defined over $\mathbb{D}^j$, i.e.,
$\boldsymbol{\Phi}_l(t_j) = \prod_{m=1}^{d_j} \phi_m^{p_m^l}(t_j)$ and 
$\boldsymbol{\Gamma}_v(t_j) = \prod_{n=1}^{d_j} \varphi_n^{p_n^v}(t_j)$. 
The total numbers of basis functions used for $X_i^{(j)}$ and $\beta^{(j)}$ are 
$\delta_j = \prod_{m=1}^{d_j} p_m^l$ and 
$s_j = \prod_{n=1}^{d_j} p_n^v$, 
where $p_m^l$ and $p_n^v$ denote the number of univariate basis functions 
used for $\phi_m$ and $\varphi_n$, respectively.

For a clearer understanding of the notation used and the methodology itself, we consider the case $k=2$, i.e., the model has two functional covariates, $X^{(1)}(t_1)$ defined over a univariate domain ($\mathbb{R}$), and $X^{(2)}(t_2)$ defined over a bivariate domain ($\mathbb{R}^2$). Then, model (1) can be reformulated as:\vspace{-0.3cm}

\begin{equation}
\label{eq:model2}
	\eta_i = g(\mu_i) = \alpha + \int\limits_{D_i^1} X_i^{(1)}(t_1)\beta^{(1)}(t_1)\text{d}t_1 + \int\limits_{D_i^2} X_i^{(2)}(t_2)\beta^{(2)}(t_2)\text{d}t_2, \quad t_1\in D_i^1 \subset \mathbb{R}, t_2\in D_i^2 \subset \mathbb{R}^2.
\end{equation}\label{functional_model_2D}

The first covariate and its corresponding coefficient are expressed through finite basis expansions as:  

\begin{eqnarray*}
	X_i^{(1)}(t_1) &=& \displaystyle\sum_{l=1}^{p_1}a^{(1)}_{il}\phi_{l}(t_1), \\
	 \beta^{(1)}(t_1) &=& \sum_{v=1}^{q_1}b^{(1)}_{v}\varphi_{v}(t_1),
\end{eqnarray*}

here $\Phi_l(t_1)=\phi_l(t_1), \; l=1,2,\ldots, p_1$ and $\Gamma_v(t_1)=\varphi_v(t_1), \; v=1,2,\ldots, q_1$ are one-dimensional B-spline basis and $\delta_1 = p_1$ and $s_1 = q_1,$ since $d_1=1$.\\

For the bivariate functional covariate, the representation becomes:

\begin{eqnarray*}
	X_i^{(2)}(\boldsymbol{t_2}) &=& X_i^{(2)}(t_{21},t_{22}) = \displaystyle \sum_{j=1}^{p_2}\sum_{k=1}^{q_2}a_{ijk}\phi_{j}(t_{21})\phi_{k}(t_{22}) = \displaystyle \sum_{l=1}^{\delta_2} \textbf{a}^{(2)}_{il}\Phi_l(\boldsymbol{t_2}), \\
	 \beta^{(2)}(\boldsymbol{t_2}) &=& \beta(t_{21}, t_{22})  =  \sum_{l=1}^{r_2}\sum_{m=1}^{c_2}\omega_{lm}\varphi_{l}(t_{21})\varphi_{m}(t_{22}) = \displaystyle \sum_{v=1}^{s_2} b^{(2)}_{v}\Gamma_v(\boldsymbol{t_2}) ,
\end{eqnarray*}

where $\delta_2 = p_2\cdot q_2$ and $s_2= r_2 \cdot c_2$. In this case, the basis $\Phi_l(\boldsymbol{t_2})$ and $\Gamma_v(\boldsymbol{t_2})$ are the tensor product bases constructed from the respective univariate bases for the functional data and the coefficient functions.

The choice of basis functions plays a crucial role in functional data analysis and is often guided by the characteristics of the data. For instance, when the data exhibit periodic behavior, a Fourier basis is typically appropriate. In contrast, when the data display strong local features and smoothness or derivative information is not essential, a wavelet basis is commonly used. In this paper, we represent the functions using  B-spline basis \citep{DeBoor2001ASplines}, which is a standard choice when the underlying functions are assumed to be smooth and when their derivatives up to a certain order are of interest.

By representing all sample functions and their corresponding functional coefficients through basis expansions, model (\ref{eq:model2}) reduces to a multivariate regression model: 

\begin{equation}		
\label{eq:model3}
 \boldsymbol{\eta} = \boldsymbol{1}\boldsymbol{A}\boldsymbol{\Psi}\boldsymbol{\theta}=\boldsymbol{B}\boldsymbol{\theta},
\end{equation}
\label{modelo multivariante}

where $\boldsymbol{A}$ is a diagonal matrix with the basis coefficients of the sample functions, and $\boldsymbol{\Psi}$ is a block column matrix comprising the inner products of the basis used in the basis representation of the sample functions and their corresponding functional coefficients:

\begin{eqnarray*}
    \boldsymbol{1} = \begin{pmatrix}
        1\\ 1 \\ \vdots \\ 1
    \end{pmatrix}, &
     \boldsymbol{A} =  \begin{pmatrix}
		 \boldsymbol{a}_1' & 0 & \ldots & 0\\
		0 & \boldsymbol{a}_2' & 0 & \ldots\\
		  \ldots & \ldots & \ldots & \ldots\\
		  0 & 0 & \ldots & \boldsymbol{a}_N' 	
	\end{pmatrix}, &    	    \boldsymbol{\Psi} =  \begin{pmatrix}
		\boldsymbol{\Psi_1}\\
		\boldsymbol{\Psi_2}\\
		\vdots\\
		\boldsymbol{\Psi_N}
	\end{pmatrix}.
	\end{eqnarray*} 


Here, the vectors $\boldsymbol{a}_i$ are constructed by stacking the basis coefficients 
associated with the basis representations of all functional covariates, as follows:

\begin{equation*}
    \boldsymbol{a}_i' = \left( \textbf{a}_i^{(1)},\textbf{a}_i^{(2)}\right) = \left( a_{i1}^{(1)}, \ldots, a_{ip_1}^{(1)}, a_{i1}^{(2)}, \ldots, a_{i\delta_2}^{(2)}\right), \quad \forall i=1, \ldots, N,
\end{equation*}

where $\boldsymbol{a}_i^{(j)}$ denotes the vector of basis coefficients corresponding 
to the $j$-th covariate for the $i$-th sample unit.

The matrices forming the block columns of the inner product matrix $\boldsymbol{\Psi}$ 
are block-diagonal, with each diagonal block corresponding to the inner product 
matrix of a specific covariate:

\begin{equation*}
         \boldsymbol{\Psi}_i =  \begin{pmatrix} \boldsymbol{\Psi}_i^{(1)} & 0 \\
         0 & \boldsymbol{\Psi}_i^{(2)}
	\end{pmatrix}.
\end{equation*}

The individual blocks $\boldsymbol{\Psi}_i^{(1)}$ and $\boldsymbol{\Psi}_i^{(2)}$ 
are defined as follows. For the first covariate, the inner product matrix is:

\begin{equation*}
    \boldsymbol{\Psi}_i^{(1)} =  \begin{pmatrix}
		\int\limits_{D_i^1} \phi'_1(t_1)\varphi_1(t_1)\text{d}t_1 & \ldots & \int\limits_{D_i^1} \phi'_1(t_1)\varphi_{q_1}(t_1)\text{d}t_1\\
		\vdots & \ddots & \vdots\\
		\int\limits_{D_i^1} \phi'_{p_1}(t_1)\varphi_1(t_1)\text{d}t_1 & \ldots & \int\limits_{D_i^1} \phi'_{p_1}(t_1)\varphi_{q_1}(t_1)\text{d}t_1
  \end{pmatrix}_{p_1\times q_1},
\end{equation*}

while for the second covariate, the corresponding matrix is:

\begin{equation*}
    \boldsymbol{\Psi}_i^{(2)} =  \begin{pmatrix}
		\int\limits_{D_i^2} \Phi'_1(t_2)\Gamma_1(t_2)\text{d}t_2 & \ldots & \int\limits_{D_i^2} \Phi'_1(t_2)\Gamma_{r_2c_2}(t_2)\text{d}t_2\\
		\vdots & \ddots & \vdots\\
		\int\limits_{D_i^2} \Phi'_{p_2q_2}(t_2)\Gamma_1(t_2)\text{d}t_2 & \ldots & \int\limits_{D_i^2} \Phi'_{p_2q_2}(t_2)\Gamma_{r_2c_2}(t_2)\text{d}t_2
  \end{pmatrix}_{p_2q_2\times r_2c_2}.
\end{equation*}

Finally, the vector of model coefficients $\boldsymbol{\theta}$ is the result of stacking all the basis coefficients associated with the functional coefficients of the model:
\begin{equation*}
    \boldsymbol{\theta} = \left(\alpha, \boldsymbol{b}^{(1)},\boldsymbol{b}^{(2)}\right)' = \left( \alpha, b_{1}^{(1)}, \ldots, b_{q_1}^{(1)}, b_{1}^{(2)}, \ldots, b_{s_2}^{(2)}\right)', \quad \forall i=1, \ldots, N.
\end{equation*}

\subsubsection{Estimation of basis coefficients for partially observed functional covariates}
\label{subsec:coef_estimation}
As described above, each functional covariate 
$X_i^{(j)}(t_j)$ is represented through a basis expansion. 
In the case of partially observed functional data, the corresponding basis coefficients 
$\boldsymbol{a_i}$ must be estimated so that the observed portions of the covariates are 
appropriately smoothed while accounting for missing observations. 
This smoothing step ensures that the estimated coefficients provide stable and consistent 
representations of the available functional information. 
For this purpose, the coefficients $\boldsymbol{a_i}$ are estimated using a penalized least squares 
approach that extends the methods proposed by \citet{Aguilera2013ComparativeData} and 
\citet{Aguilera-Morillo2017} to accommodate partially observed functional data. 
Specifically, we include a discrete penalty term based on second-order differences between 
adjacent B-spline coefficients \citep{Eilers1996FlexiblePenalties}. 
Adapting this standard penalization approach to partially observed functional data, however, 
requires several modifications. In particular, the basis used to represent the sampled 
functional covariate $X_i^{(j)}(t_j)$ is first evaluated over its entire domain 
$\mathbb{D}^j$, as if the function were fully observed. 
The resulting B-spline design matrix is then adjusted by multiplying it by a weighted 
identity matrix that assigns zeros to entries corresponding to missing observations. 
This modified design matrix is subsequently used in the penalized least squares regression 
to estimate the basis coefficients $\boldsymbol{a_i}$. 
The following algorithm summarizes this procedure for the specific case of a 
two-dimensional functional coefficient defined over the domain 
$\mathbb{D} = [0,1] \times [0,1]$.

\begin{algorithm}[H]
\caption{Penalized least squares estimation for partially observed 2D functional data}
\textbf{Input:} $\{X_i(t_1,t_2)\}_{i=1}^N$ - Collection of partially observed 2D functions\\
\phantom{\textbf{Input:}} $\mathbb{D} = [0,1] \times [0,1]$ - Complete domain\\
\phantom{\textbf{Input:}} $\mathcal{M}_i$ - Missing observation indices for function $i$\\
\phantom{\textbf{Input:}} $p_2, q_2$ - Number of B-spline basis functions for each dimension\\
\textbf{Output:} $\mathbf{A}$ - Matrix of estimated basis coefficients\newline
\textbf{Steps:}\newline
1: \textbf{Initialize B-spline basis systems:}
\begin{enumerate}
    \item Construct B-spline basis $\boldsymbol{\Phi}(t_1,t_2) = \boldsymbol{\phi}(t_1) \otimes \boldsymbol{\phi}(t_2)$ with $p_2 \times q_2$ functions
\end{enumerate}
2: \textbf{For each function} $i = 1, \ldots, N$:\newline
\phantom{2:} a) Identify missing value positions:\newline
\phantom{a)} \textbf{For} $(r,c) \in \{1,\ldots,\text{nrow}(X_i)\} \times \{1,\ldots,\text{ncol}(X_i)\}$:\newline
\phantom{a)} \phantom{\textbf{For}} \textbf{If} $X_i[r,c] = \text{NA}$: $\mathcal{I}_i \leftarrow \mathcal{I}_i \cup \{(\text{nrow}(X_i) \times (c-1)) + r\}$\newline
\phantom{2:} b) Create weight matrix: $\mathbf{I}_i = \text{diag}(\text{nrow}(\boldsymbol{\Phi}))$ and set $\mathbf{I}_i[\text{NA\_ind}, :] = \mathbf{0}$\newline
\phantom{2:} c) Modify design matrix: $\boldsymbol{\Phi}_i^* = \mathbf{I}_i \boldsymbol{\Phi}$\newline
\phantom{2:} d) Estimate coefficients as described in the mentioned literature using the new design matrix $\boldsymbol{\Phi}_i^*$\newline
3: \textbf{Return:} Set of basis coefficients: $\{\hat{\boldsymbol{a}}_i\}_{i=1}^N$ 
\end{algorithm}

The estimated coefficients serve as the inputs for the regression model (\ref{eq:model3}), which is fitted using a mixed model representation presented in the next section.
\subsection{Model estimation through a mixed model representation}
\label{2.4}

The multivariate regression model (\ref{eq:model3}) falls into the category of generalized linear models and therefore the maximum likelihood method is used to estimate the model parameters.

Since the functional coefficient has been represented using B-spline basis, the smoothness of the resulting estimated coefficient is determined by the basis dimension. To avoid the problem of choosing the optimal number of basis functions we follow the penalized likelihood approach by \cite{Eilers1996FlexiblePenalties} defined as:

$$L_p(\boldsymbol{\theta},\boldsymbol{y}) = L(\boldsymbol{\theta},\boldsymbol{y}) - \frac{1}{2} \boldsymbol{\theta' P\theta},$$

where $L(\boldsymbol{\theta},\boldsymbol{y})$ is the likelihood of $\boldsymbol{Y}$ and $ \boldsymbol{P}$ is the penalty term that controls the smoothness of the function via one or more smoothing parameters.\\






The penalty matrix $\boldsymbol{P}$ must account for the unpenalized intercept term and the different penalization structures associated with the one-dimensional and two-dimensional functional coefficients. For the two-dimensional covariate, an anisotropic penalty is employed to control smoothness separately along each dimension. The resulting penalty matrix is:

\begin{equation}
\boldsymbol{P} = \text{blockdiag}\{0, \boldsymbol{P}^{(1)}, \boldsymbol{P}^{(2)}\},
\end{equation}

where the leading zero ensures that the intercept term is not being penalized and $\boldsymbol{P}^{(1)}, \boldsymbol{P}^{(2)}$ are the specific penalty matrices for each covariate with:

\begin{eqnarray*}
\boldsymbol{P}^{(1)} &=& \lambda_{t_1}\boldsymbol{D}_{t_1}'\boldsymbol{D}_{t_1},\\
\boldsymbol{P}^{(2)} &=& \lambda_{t_{21}}(\boldsymbol{I}_r \otimes \boldsymbol{D}_{t_{21}}' \boldsymbol{D}_{t_{21}}')+\lambda_{t_{22}}(\boldsymbol{I}_q \otimes \boldsymbol{D}_{t_{22}}' \boldsymbol{D}_{t_{22}}),
\end{eqnarray*}

with $\boldsymbol{D}_{t_1}$, $\boldsymbol{D}_{t_{21}}$ and $\boldsymbol{D}_{t_{22}}$ are second-difference matrices, and $\otimes$ denotes the Kronecker product. 

 
Finally, we use the mixed model reparametrization of a penalized spline to estimate the parameters of the model. This approach enables the simultaneous estimation 
of all model parameters, including the smoothing parameters. A brief overview of the 
reparameterization is provided below to clarify the methodology used. For a comprehensive discussion of the mixed model representation of penalized splines 
in the context of fully observed functional data,  see \cite{Lee2010SmoothingData}. 

Our goal is to express the penalized spline formulation in a mixed model framework, transforming
 
\begin{equation}
    \boldsymbol{\eta}=\boldsymbol{B}\boldsymbol{\theta} \Rightarrow \boldsymbol{X}\boldsymbol{\nu} + \boldsymbol{Z}\boldsymbol{\delta},\quad \boldsymbol{\delta}\sim N(0,\boldsymbol{G}),
\end{equation}\label{mixed model}
where $\boldsymbol{X}$ and $\boldsymbol{Z}$ denote the model matrices, 
$\boldsymbol{\nu}$ and $\boldsymbol{\delta}$ the fixed and random effects, respectively, 
and $\boldsymbol{G}$ the variance–covariance matrix of the random effects. 
The matrix $\boldsymbol{G}$ depends on the variance components 
$\tau_{t_1}^2$, $\tau_{t_{21}}^2$, and $\tau_{t_{22}}^2$. 

This reparameterization is achieved using a transformation matrix $\boldsymbol{T}$ 
based on the Singular Value Decomposition (SVD) of the product of difference matrices 
$\boldsymbol{D}_i'\boldsymbol{D}_i$. Let
\begin{equation*}
    \boldsymbol{D}_i'\boldsymbol{D}_i = 
    \left[\boldsymbol{U}_{in} \mid \boldsymbol{U}_{is}\right]
    \begin{bmatrix}
        \boldsymbol{0}_2 & \\
        & \tilde{\boldsymbol{\Sigma}}_i
    \end{bmatrix}
    \begin{bmatrix}
        \boldsymbol{U}_{in}'\\[2pt]
        \boldsymbol{U}_{is}'
    \end{bmatrix},
\end{equation*}
be the SVD factorization of $\boldsymbol{D}_i'\boldsymbol{D}_i$ for 
$i \in \{t_1, t_{21}, t_{22}\}$, where $\boldsymbol{U}_{in}$ and $\boldsymbol{U}_{is}$ 
are the eigenvector matrices associated with the zero and nonzero eigenvalues, respectively. 
The transformation matrix $\boldsymbol{T}$ is then defined as
\begin{equation*}
    \boldsymbol{T} =
    \begin{bmatrix}
        1 & \boldsymbol{0} & \boldsymbol{0} \\
        \boldsymbol{0} & \boldsymbol{T}_n & \boldsymbol{T}_s
    \end{bmatrix},
\end{equation*}
where $\boldsymbol{T}_n = \mathrm{blockdiag}\{T_n^{(1)}, T_n^{(2)}\}$ and 
$\boldsymbol{T}_s = \mathrm{blockdiag}\{T_s^{(1)}, T_s^{(2)}\}$.

Here, the matrices $\boldsymbol{T}_n^{(j)}$ and $\boldsymbol{T}_s^{(j)}$ correspond to 
the non-penalized and penalized coefficients, respectively, for each functional covariate. 
Their computation differs between the one-dimensional covariate ($j = 1$) and the 
two-dimensional covariate ($j = 2$), as follows:
\begin{align*}
    \boldsymbol{T}_n^{(1)} &= \boldsymbol{U}_{t_{1}n}, \\
    \boldsymbol{T}_s^{(1)} &= \boldsymbol{U}_{t_{1}s}, \\
    T_n^{(2)} &= \boldsymbol{U}_{t_{22}n} \otimes \boldsymbol{U}_{t_{21}n}, \\
    T_s^{(2)} &= 
    \left[
        \boldsymbol{U}_{t_{22}s} \otimes \boldsymbol{U}_{t_{21}n} \;\middle|\;
        \boldsymbol{U}_{t_{22}n} \otimes \boldsymbol{U}_{t_{21}s} \;\middle|\;
        \boldsymbol{U}_{t_{22}s} \otimes \boldsymbol{U}_{t_{21}s}
    \right].
\end{align*}

Among the possible formulations of the transformation matrix $\boldsymbol{T}$, 
the one proposed in this work is particularly advantageous because the orthogonality of 
$\boldsymbol{T}$ ensures that the original parameter estimates 
$\boldsymbol{\widehat{\theta}}$ and hence, the functional coefficients can be directly 
recovered from the mixed model estimates. 

Using this transformation matrix, model~\ref{eq:model3} can be reparameterized as:
\[
\boldsymbol{\eta} = \boldsymbol{B}\boldsymbol{\theta} 
= \boldsymbol{B}\boldsymbol{T}\boldsymbol{T}'\boldsymbol{\theta} 
= \boldsymbol{X}\boldsymbol{\nu} + \boldsymbol{Z}\boldsymbol{\delta},
\]
where $\boldsymbol{B}\boldsymbol{T} = [\,\boldsymbol{B}\boldsymbol{T_n} \mid \boldsymbol{B}\boldsymbol{T_s}\,] 
= [\,\boldsymbol{X} \mid \boldsymbol{Z}\,]$, and 
$\boldsymbol{T}'\boldsymbol{\theta} = \boldsymbol{\omega}$ with 
$\boldsymbol{\omega}' = (\boldsymbol{\nu}', \boldsymbol{\delta}')$. 
The variance–covariance matrix of the random effects, $\boldsymbol{G}$, is derived by 
applying the transformation to the penalty matrix, yielding 
$\boldsymbol{G}^{-1} = \boldsymbol{T}'\boldsymbol{P}\boldsymbol{T}$, where
\begin{equation}
\boldsymbol{G}^{-1} = 
\begin{pmatrix}
	\dfrac{1}{\tau_{t_1}^2}\tilde{\boldsymbol{\Sigma}}_{t_1} & & & \\
	& \dfrac{1}{\tau_{t_{22}}^2}\tilde{\boldsymbol{\Sigma}}_{t_{22}} \otimes \boldsymbol{I}_2 & & \\
	& & \dfrac{1}{\tau_{t_{21}}^2}\boldsymbol{I}_2 \otimes \tilde{\boldsymbol{\Sigma}}_{t_{21}} & \\
	& & & 
	\dfrac{1}{\tau_{t_{22}}^2}\tilde{\boldsymbol{\Sigma}}_{t_{22}} \otimes \boldsymbol{I}_{q-2} 
	+ \dfrac{1}{\tau_{t_{21}}^2}\boldsymbol{I}_{r-2} \otimes \tilde{\boldsymbol{\Sigma}}_{t_{21}}
\end{pmatrix}.
\end{equation}

The variance components are related to the smoothing parameters through 
$\tau_{t_1}^2 = 1/\lambda_{t_1}$, 
$\tau_{t_{21}}^2 = 1/\lambda_{t_{21}}$, and 
$\tau_{t_{22}}^2 = 1/\lambda_{t_{22}}$ \citep{Brumback1999}. 
This mixed model representation not only provides a unified framework for estimating 
the model parameters and smoothing components simultaneously, but also facilitates the 
recovery of the functional coefficients in their original space. 
Finally, the mixed model coefficients are estimated using the SOP algorithm 
\citep{Rodriguez-Alvarez2019OnSmoothing}.


\section{Simulation study} \label{section sim}

To evaluate the performance of the proposed model, two simulation studies have been conducted in this section. The first study comprises 24 scenarios involving two one-dimensional 
functional covariates, while the second study focuses on a single two-dimensional 
functional covariate across 18 scenarios. In both studies, two types of response 
variables were considered: Normal and Binomial. The different scenarios arise from 
combinations of varying proportions of partially observed functional data, gap sizes, 
and choices of the underlying functional coefficients.

In both simulation settings, we benchmarked the proposed method against an imputation-based 
strategy in which missing segments of the functional data were first filled and then the 
completed data were used to fit the model. For this strategy, we adopted the imputation 
method proposed by \citet{Kraus2015ComponentsData}, which is designed for one-dimensional functional data. 
Since this approach is not directly applicable to two-dimensional functional data, we 
extended it by vectorizing the functional surfaces into curves, thereby enabling its use 
in the second simulation study.

However, the practical implementation of the method by \cite{Kraus2015ComponentsData} revealed several 
limitations that depend on the characteristics of the dataset. In the one-dimensional 
simulation experiments, the method failed when more than 80\% of the curves contained 
missing segments, requiring at least 20\% of the curves to be fully observed for reliable 
performance. Furthermore, the method exhibited sensitivity to gap size, particularly when 
the proportion of incomplete curves was high. These constraints became even more restrictive 
in the two-dimensional setting, where Kraus's method required missing regions to be limited 
to approximately 60\% of each surface. In addition, vectorizing a $20 \times 20$ surface into 
a 400-dimensional vector substantially increased computational complexity. Despite these 
empirically observed limitations, this method was chosen as a benchmark because it is among 
the few existing approaches that can be adapted to handle two-dimensional functional data.

\subsection{Simulation Study 1: additive one dimensional linear predictors}

For simplicity, we consider models including only functional covariates, although the 
proposed framework can easily accommodate additional non-functional predictors. 
We simulated 100 datasets of sample size $N = 100$ for each combination of the parameters 
described below, resulting in 24 distinct scenarios:

\begin{itemize}
    \item \textbf{Outcome types}: Two types of outcomes were considered, continuous and binary data, both generated with the following linear predictor:

\begin{equation*}
    \eta_i = \alpha + \displaystyle \int_{0}^{1} X_{i1}(t) \beta_1(t)\text{d}t + \int_{0}^{1} X_{i2}(t) \beta_2(t)\text{d}t, \;\; t\in [0,1].
		\end{equation*}

For continuous outcomes, $Y_i = \eta_i + \epsilon_i$ with 
    $\epsilon_i \sim N(0, \delta_i)$, where $\delta_i$ is chosen such that the true 
    model’s squared correlation coefficient ($R^2$) equals 0.95. 
    For binary outcomes, $Y_i \sim \mathrm{Bernoulli}(p_i)$ with 
    $p_i = \frac{\exp(\eta_i)}{1 + \exp(\eta_i)}$. 
    In all cases, the intercept was fixed at $\alpha = 0.5$.\\

\item \textbf{Percentage of curves with missing observations}: 20\%, 40\%, 60\% and 80\%.\\

\item \textbf{Percentage of missing observations (gap size)}: 5\%, 10\%, 15\%.\\

\item \textbf{Functional covariate generation}:
\begin{eqnarray*}
    X_{i1}(t) &=& u_{i1} + u_{i2} \cdot t + \sum_{k=1}^{10} \left[ v_{i1k} \sin\left(\frac{2\pi k}{10} t\right) + v_{i2k} \cos\left(\frac{2\pi k}{10} t\right) \right] + \epsilon_{i1}(t), \\
    X_{i2}(t) &=& u_{i1}' + u_{i2}' \cdot t + \sum_{k=1}^{10} \left[ v_{i1k}' \sin\left(\frac{2\pi (k+0.5)}{10} t + \frac{\pi}{4}\right) + v_{i2k}' \cos\left(\frac{2\pi (k+0.5)}{10} t - \frac{\pi}{4}\right) \right] + \epsilon_{i2}(t),
\end{eqnarray*}
 where $t = \{0, t_2, \ldots, t_{99}, 1\}$ are 100 equidistant points in $[0,1]$. 
    The random coefficients were generated as follows: 
    $u_{i1} \sim N(0, 5)$, $u_{i2} \sim N(0, 0.2)$, 
    $u_{i1}' \sim N(2, 3)$, $u_{i2}' \sim N(-0.1, 0.15)$, 
    $v_{i1k}, v_{i2k} \sim N(0, 1)$, 
    $v_{i1k}' \sim N(-0.5, 0.8)$, and 
    $v_{i2k}' \sim N(0.5, 0.8)$ for $k = 1, \ldots, 10$. 
    The error terms were generated as $\epsilon_{il}(t) \sim N(0, \sigma_{il})$, with 
    $\sigma_{il} = 0.5\sqrt{\mathrm{Var}(X_{il}(t))}$ for $l \in \{1,2\}$. 
    This ensures that the noise variance is one quarter of the corresponding covariate 
    variance and remains consistent across curves.\\

\item \textbf{Functional coefficient configurations}:
\begin{eqnarray*}
    \beta_1(t) &= 0.05 \sin\left(\frac{\pi t}{5}\right), \\
    \beta_2(t) &= 0.05 \left(\frac{t}{2.5}\right)^2.
\end{eqnarray*}
\end{itemize}

\subsubsection{Performance criteria}

We evaluate the proposed methodology along two key dimensions: (i) predictive performance and (ii) its ability to recover the true functional parameter. Model performance was assessed differently depending on the type of response variable.

\begin{itemize}
    \item \textbf{Binary outcomes:} Predictive accuracy was evaluated using the 
    misclassification error and the area under the receiver operating characteristic curve (AUC).\\[3pt]

    \item \textbf{Continuous outcomes:} Predictive accuracy was measured using the 
    root mean square error (RMSE), defined as
    \begin{equation*}
        RMSE = \sqrt{\frac{1}{N}\sum_{i=1}^{N} (Y_i - \hat{Y}_i)^2}.
    \end{equation*}
\end{itemize}

The second evaluation criterion concerns the model’s ability to recover the true 
functional coefficients, $\beta_{1}(t)$ and $\beta_{2}(t)$. 
This was quantified using the integrated mean square error (IMSE), defined as:

\begin{equation*}
	IMSE_l = \displaystyle \int_{0}^{1} \left(\beta_{l}(t) - \hat{\beta}_{l}(t)\right)^2\text{d}t, \, l=\{1,2\},
\end{equation*}

where \( \hat{\beta}_{l}(t) \) denotes the estimated functional coefficient.

\subsubsection{Results}\label{1D results}

This section presents the results of the one-dimensional simulation study. The outcomes are summarized in tables and visualized using violin plots. Tables report mean values with standard deviations in parentheses for interpretability. 

In the following discussion, we denote by \textbf{POFRM} the \emph{Partially Observed Functional Regression Model} proposed, which directly estimates model parameters from the partially observed functional data without any imputation step. Conversely, \textbf{POFRM-I} refers to the imputation-based version of the model, where missing segments are first reconstructed using the method proposed by \citet{Kraus2015ComponentsData} before model estimation.

Tables~\ref{Beta_1_Normal}--\ref{Y_Normal} summarize the results for the Normal response scenarios, reporting the integrated mean squared error (IMSE) of the functional coefficient estimates and the root mean squared error (RMSE) of the response variable. Tables~\ref{Beta_1_Binomial}--\ref{Y_Binomial} present analogous results for the Binomial response, including the IMSEs for the functional coefficients and both the area under the curve (AUC) and misclassification error for predictive performance. Selected scenarios are illustrated in violin plots to visualize the distribution and variability of the estimation errors. All additional figures are provided in the supplementary material.
\\

\vspace{3mm}
\noindent\textbf{Normal response scenarios}

\vspace{2mm}

\begin{table}[htbp]
\centering
\small
\begin{tabular}{|l|l|c|c|c|}
\hline
Curves with gaps & Method & \multicolumn{3}{c|}{Gap size} \\
& & 5\% & 10\% & 15\% \\
\hline
\multirow{2}{*}{20\%} & POFRM & \textbf{0.010 (0.003)} & \textbf{0.010 (0.003)} & \textbf{0.011 (0.005)} \\
 & POFRM-I & 0.012 (0.004) & 0.015 (0.006) & 0.019 (0.005) \\
\hline
\multirow{2}{*}{40\%} & POFRM & \textbf{0.010 (0.003)} & \textbf{0.011 (0.004)} & \textbf{0.011 (0.005)} \\
 & POFRM-I & 0.013 (0.005) & 0.018 (0.006) & 0.023 (0.001) \\
\hline
\multirow{2}{*}{60\%} & POFRM & \textbf{0.010 (0.004)} & \textbf{0.012 (0.005)} & \textbf{0.013 (0.002)} \\
 & POFRM-I & 0.014 (0.006) & 0.021 (0.005) & 0.024 (0.005) \\
\hline
\multirow{2}{*}{80\%} & POFRM & \textbf{0.011 (0.004)} & \textbf{0.013 (0.005)} & \textbf{0.015 (0.006)} \\
 & POFRM-I & 0.015 (0.006) & 0.023 (0.003) & 0.025 (0.002) \\
\hline
\end{tabular}
\caption{Mean and standard deviation (in parentheses) of 100 measures of the IMSE for $\beta_1(t)$ for the scenarios where the response variable follows a Normal distribution in the 1D simulation study.}
\label{Beta_1_Normal}
\end{table}

\begin{table}[htbp]
\centering
\small
\begin{tabular}{|l|l|c|c|c|}
\hline
Curves with gaps & Method & \multicolumn{3}{c|}{Gap size} \\
& & 5\% & 10\% & 15\% \\
\hline
\multirow{2}{*}{20\%} & POFRM & \textbf{0.033 (0.009)} & \textbf{0.033 (0.008)} & \textbf{0.034 (0.009)} \\
 & POFRM-I & 0.033 (0.010) & 0.033 (0.01) & 0.18 (0.12) \\
\hline
\multirow{2}{*}{40\%} & POFRM & \textbf{0.033 (0.008)} & \textbf{0.032 (0.008)} & \textbf{0.032 (0.009)} \\
 & POFRM-I & 0.036 (0.012) & 0.079 (0.047) & 0.2 (0.12) \\
\hline
\multirow{2}{*}{60\%} & POFRM & \textbf{0.033 (0.008)} & \textbf{0.032 (0.009)} & \textbf{0.032 (0.009)} \\
 & POFRM-I & 0.038 (0.013) & 0.081 (0.028) & 0.18 (0.085) \\
\hline
\multirow{2}{*}{80\%} & POFRM & \textbf{0.032 (0.008)} & \textbf{0.031 (0.009)} & \textbf{0.030 (0.011)} \\
 & POFRM-I & 0.040 (0.012) & 0.084 (0.018) & 0.144 (0.040) \\
\hline
\end{tabular}
\caption{Mean and standard deviation (in parentheses) of 100 measures of the IMSE for $\beta_2(t)$ for the scenarios where the response variable follows a Normal distribution in the 1D simulation study.}
\label{Beta_2_Normal}
\end{table}

\begin{table}[htbp]
\centering
\small
\begin{tabular}{|l|l|c|c|c|}
\hline
Curves with gaps & Method & \multicolumn{3}{c|}{Gap size} \\
& & 5\% & 10\% & 15\% \\
\hline
\multirow{2}{*}{20\%} & POFRM & \textbf{0.51 (0.033)} & \textbf{0.52 (0.03)} & \textbf{0.53 (0.035)} \\
 & POFRM-I & 0.57 (0.037) & 0.68 (0.051) & 0.85 (0.10) \\
\hline
\multirow{2}{*}{40\%} & POFRM & \textbf{0.51 (0.032)} & \textbf{0.53 (0.03)} & \textbf{0.54 (0.037)} \\
 & POFRM-I & 0.63 (0.039) & 0.84 (0.068) & 1.12 (0.13) \\
\hline
\multirow{2}{*}{60\%} & POFRM & \textbf{0.52 (0.031)} & \textbf{0.54 (0.033)} & \textbf{0.57 (0.039)} \\
 & POFRM-I & 0.70 (0.037) & 1 (0.068) & 1.36 (0.12) \\
\hline
\multirow{2}{*}{80\%} & POFRM & \textbf{0.52 (0.034)} & \textbf{0.56 (0.037)} & \textbf{0.606 (0.043)} \\
 & POFRM-I & 0.76 (0.04) & 1.16 (0.070) & 1.6 (0.11) \\
\hline
\end{tabular}
\caption{Mean and standard deviation (in parentheses) of 100 measures of the RMSE for the scenarios where the response variable follows a Normal distribution in the 1D simulation study.}
\label{Y_Normal}
\end{table}

The results in Tables \ref{Beta_1_Normal}, \ref{Beta_2_Normal}, and \ref{Y_Normal} demonstrate that the proposed POFRM method clearly outperforms the imputation-based approach (POFRM-I) across all scenarios and evaluation metrics. Both functional coefficient estimates, $\beta_1(t)$ and $\beta_2(t)$, exhibit smaller IMSE values under POFRM, with the improvement becoming more pronounced as the proportion of incomplete curves and the gap size increase.

In particular, while POFRM-I’s accuracy deteriorates noticeably when 60–80

The superiority of POFRM extends to predictive performance as well. The RMSE results in Table \ref{Y_Normal} indicate that the proposed method achieves systematically lower prediction errors than POFRM-I across all levels of missingness. Even under the most challenging conditions—large gaps and 80$\% $ incomplete curves—the maximum RMSE obtained by POFRM remains below the minimum error reached by the imputation-based alternative.

\begin{figure}[H]
    \centering
\includegraphics[width=0.75\linewidth]{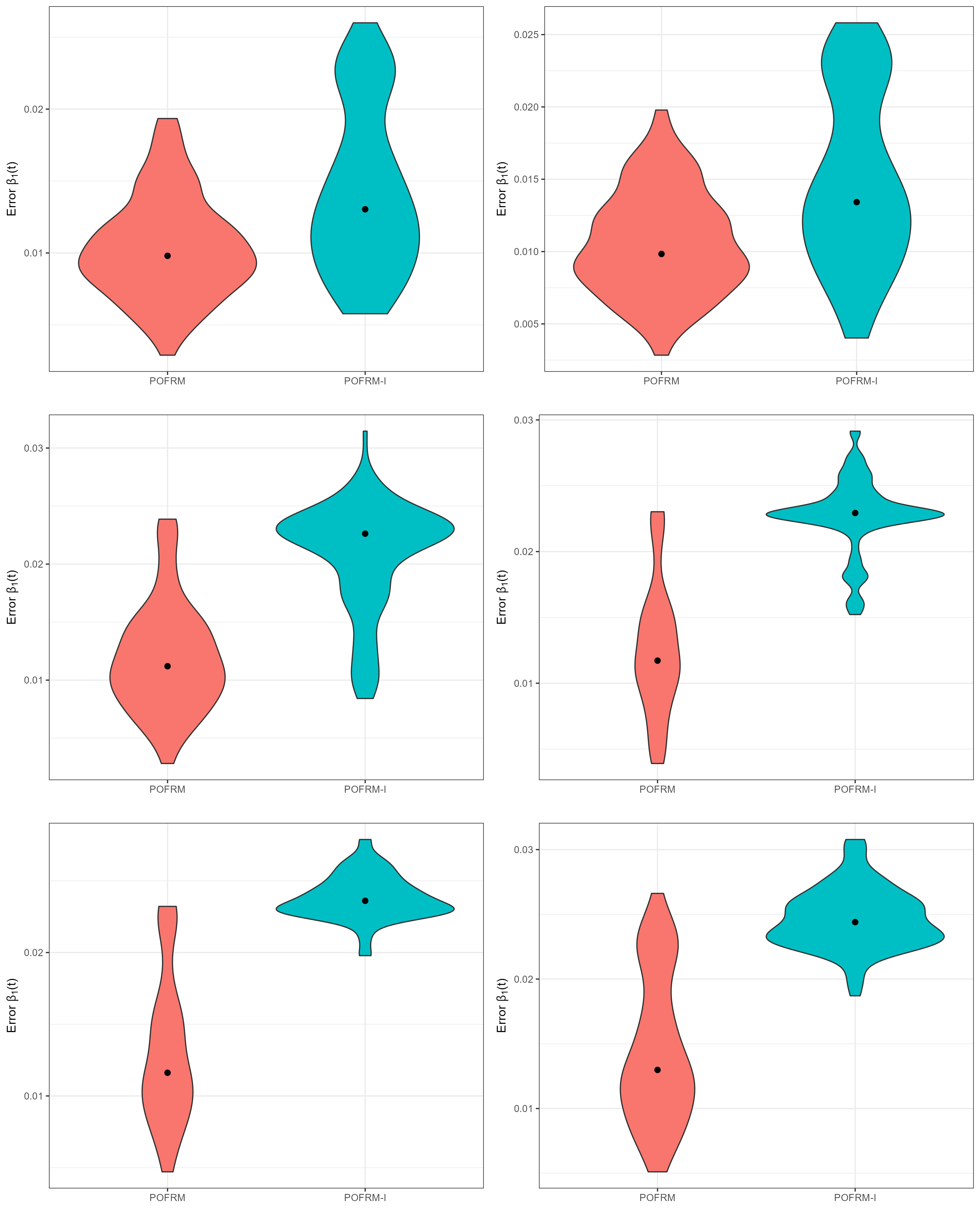}
    \caption{IMSE for the functional coefficient $\beta_1(t)$ for the 1D simulation study with Gaussian response. The left column corresponds to scenarios with 60\% of curves containing gaps, and the right column to 80\%. Rows correspond (top to bottom) to gap sizes of 5\%, 10\%, and 15\%.}
    \label{Beta_1_fig}
\end{figure}

\begin{figure}[H]
    \centering
\includegraphics[width=0.75\linewidth]{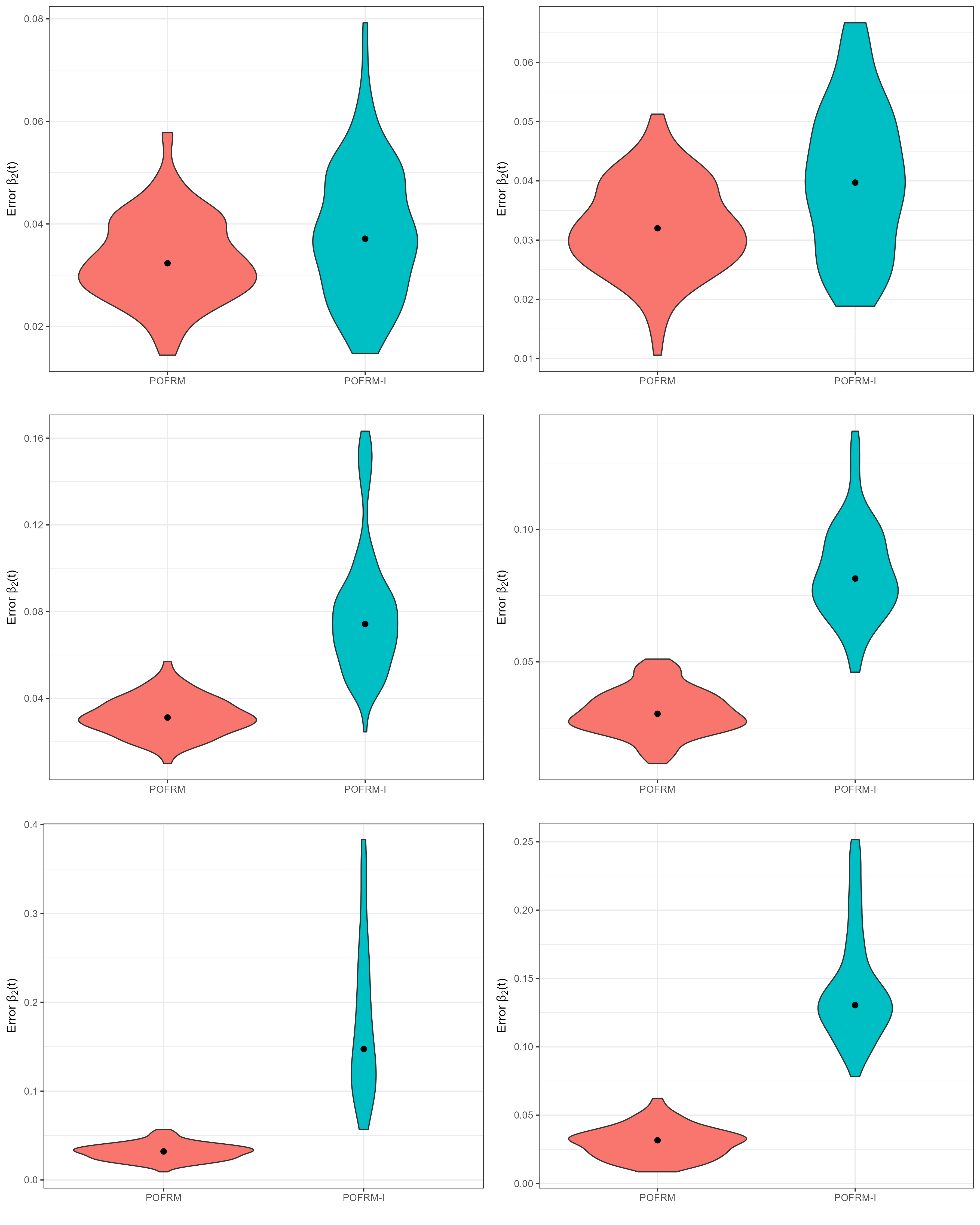}
    \caption{IMSE for the functional coefficient $\beta_2(t)$ for the 1D simulation study with Gaussian response. The left column corresponds to scenarios with 60\% of curves containing gaps, and the right column to 80\%. Rows correspond (top to bottom) to gap sizes of 5\%, 10\%, and 15\%.}
    \label{Beta_2_fig}
\end{figure}

\begin{figure}[H]
    \centering
\includegraphics[width=0.75\linewidth]{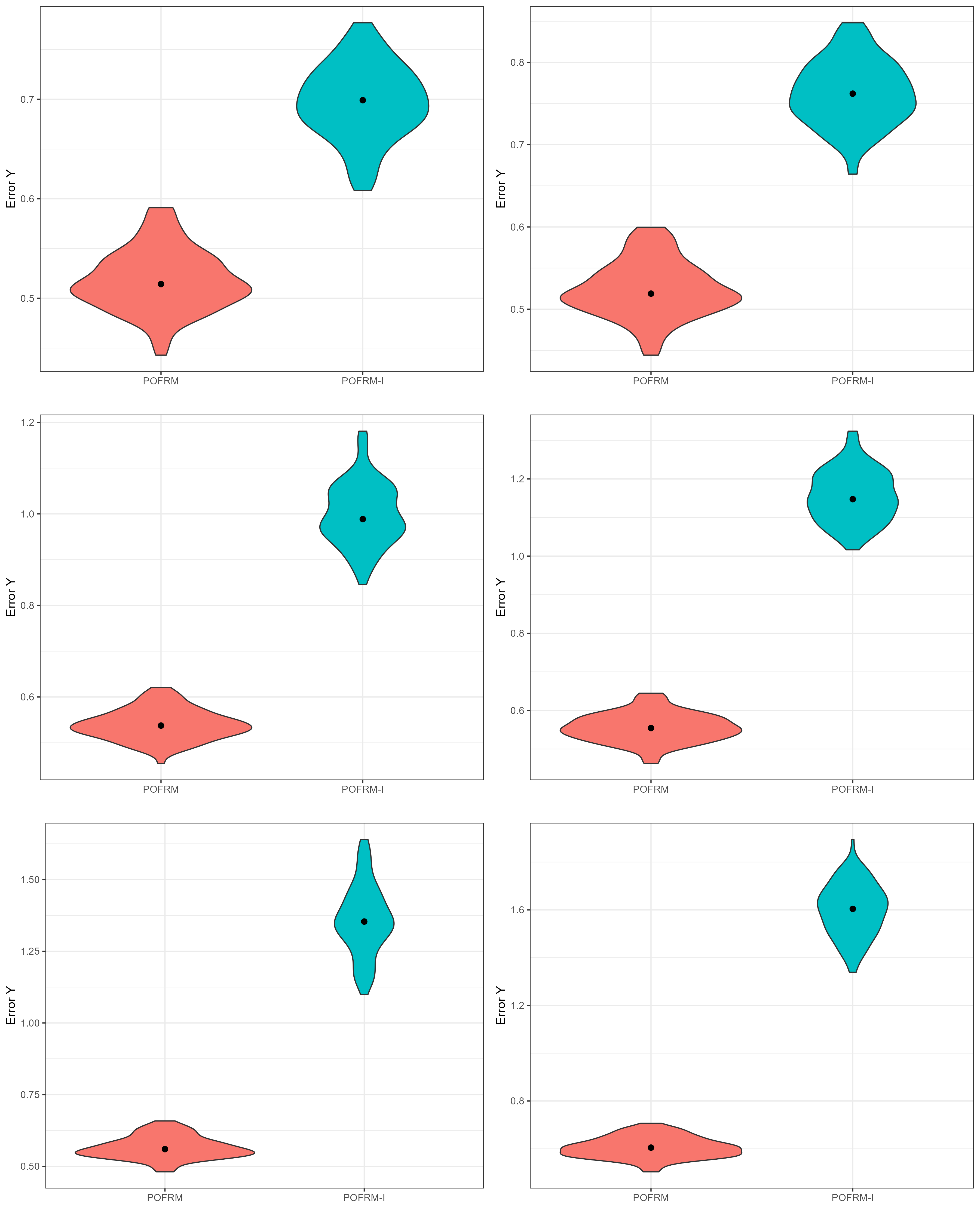}
    \caption{RMSE of the response variable for the 1D simulation study with Gaussian response. The left column corresponds to scenarios with 60\% of curves containing gaps, and the right column to 80\%. Rows correspond (top to bottom) to gap sizes of 5\%, 10\%, and 15\%.}
    \label{Y_Normal_fig}
\end{figure}

\vspace{3mm}
\noindent\textbf{Binomial response scenarios}

\vspace{2mm}

\begin{table}[htbp]
\centering
\small
\begin{tabular}{|l|l|c|c|c|}
\hline
Curves with gaps & Method & \multicolumn{3}{c|}{Gap size} \\
& & 5\% & 10\% & 15\% \\
\hline
\multirow{2}{*}{20\%} & POFRM & \textbf{0.058 (0.032)} & 0.060 (0.034) & \textbf{0.053 (0.025)} \\
 & POFRM-I & 0.061 (0.034) & \textbf{0.059 (0.034)} & 0.055 (0.025) \\
\hline
\multirow{2}{*}{40\%} & POFRM & \textbf{0.058 (0.031)} & 0.062 (0.036) & \textbf{0.048 (0.021)} \\
 & POFRM-I & 0.061 (0.035) & \textbf{0.061 (0.034)} & 0.053 (0.024) \\
\hline
\multirow{2}{*}{60\%} & POFRM & \textbf{0.057 (0.031)} & \textbf{0.059 (0.031)} & \textbf{0.045 (0.019)} \\
 & POFRM-I & 0.061 (0.035) & 0.060 (0.028) & 0.051 (0.023) \\
\hline
\multirow{2}{*}{80\%} & POFRM & \textbf{0.057 (0.034)} & \textbf{0.061 (0.035)} & \textbf{0.049 (0.025)} \\
 & POFRM-I & 0.061 (0.036) & 0.062 (0.028) & 0.055 (0.027) \\
\hline
\end{tabular}
\caption{Mean and standard deviation (in parentheses) of 100 measures of the IMSE for $\beta_1(t)$ for the scenarios where the response variable follows a Binomial distribution in the 1D simulation study.}
\label{Beta_1_Binomial}
\end{table}

\begin{table}[htbp]
\centering
\small
\begin{tabular}{|l|l|c|c|c|}
\hline
Curves with gaps & Method & \multicolumn{3}{c|}{Gap size} \\
& & 5\% & 10\% & 15\% \\
\hline
\multirow{2}{*}{20\%} & POFRM & \textbf{0.143 (0.079)} & \textbf{0.148 (0.086)} & \textbf{0.143 (0.068)} \\
 & POFRM-I & 0.145 (0.082) & 0.153 (0.087) & 0.146 (0.083) \\
\hline
\multirow{2}{*}{40\%} & POFRM & \textbf{0.135 (0.067)} & 0.156 (0.090) & 0.150 (0.077) \\
 & POFRM-I & 0.147 (0.084) & \textbf{0.143 (0.069)} & \textbf{0.148 (0.083)} \\
\hline
\multirow{2}{*}{60\%} & POFRM & \textbf{0.137 (0.066)} & \textbf{0.152 (0.087)} & \textbf{0.147 (0.074)} \\
 & POFRM-I & 0.146 (0.082) & 0.154 (0.083) & 0.148 (0.083) \\
\hline
\multirow{2}{*}{80\%} & POFRM & 0.147 (0.083) & \textbf{0.148 (0.074)} & 0.145 (0.073) \\
 & POFRM-I & \textbf{0.137 (0.067)} & 0.150 (0.084) & \textbf{0.144 (0.077)} \\
\hline
\end{tabular}
\caption{Mean and standard deviation (in parentheses) of 100 measures of the IMSE for $\beta_2(t)$ for the scenarios where the response variable follows a Binomial distribution in the 1D simulation study.}
\label{Beta_2_Binomial}
\end{table}

\begin{table}[htbp]
\centering
\small
\begin{tabular}{|l|l|c|c|c|}
\hline
Curves with gaps & Method & \multicolumn{3}{c|}{Gap size} \\
 & & 5\% & 10\% & 15\% \\ \hline
\multirow{2}{*}{20\%} & POFRM & \textbf{0.9902} (0.0047) & \textbf{0.99022} (0.00466) & \textbf{0.99} (0.0046) \\
 & POFRM-I & 0.99 (0.0046) & 0.99017 (0.005) & 0.989 (0.00455) \\
\hline
\multirow{2}{*}{40\%} & POFRM & \textbf{0.99} (0.00469) & 0.9899 (0.005) & 0.9894 (0.00477) \\
 & POFRM-I & 0.989 (0.00462) & \textbf{0.99} (0.0047) & \textbf{0.9899} (0.00472) \\
\hline
\multirow{2}{*}{60\%} & POFRM & \textbf{0.99022} (0.0046) & \textbf{0.99018} (0.0048) & \textbf{0.989} (0.0047) \\
 & POFRM-I & 0.98983 (0.0046) & 0.98986 (0.0052) & 0.988 (0.0048) \\
\hline
\multirow{2}{*}{80\%} & POFRM & 0.9895 (0.00485) & 0.989 (0.0047) & 0.988 (0.005) \\
 & POFRM-I & \textbf{0.9899} (0.0048) & \textbf{0.99} (0.00476) & \textbf{0.989} (0.0048) \\
\hline
\end{tabular}
\caption{Mean and standard deviation (in parentheses) of AUC values for Binomial response in the 1D simulation study.}
\label{AUC_Binomial_1D}
\end{table}

\begin{table}[htbp]
\centering
\small
\begin{tabular}{|l|l|c|c|c|}
\hline
Curves with gaps & Method & \multicolumn{3}{c|}{Gap size} \\
& & 5\% & 10\% & 15\% \\
\hline
\multirow{2}{*}{20\%} & POFRM & \textbf{9.85 (3.47)} & \textbf{9.66 (3.42)} & \textbf{9.80 (2.975)} \\
 & POFRM-I & 9.87 (3.36) & 9.78 (3.40) & 9.87 (2.99) \\
\hline
\multirow{2}{*}{40\%} & POFRM & \textbf{9.71 (3.22)} & \textbf{9.61 (3.40)} & \textbf{9.92 (3.23)} \\
 & POFRM-I & 9.75 (3.23) & 9.94 (3.34) & 10.43 (3.27) \\
\hline
\multirow{2}{*}{60\%} & POFRM & \textbf{9.67 (2.98)} & \textbf{9.65 (3.33)} & \textbf{9.84 (3)} \\
 & POFRM-I & 9.84 (3.3) & 10.19 (3.64) & 10.64 (3.37) \\
\hline
\multirow{2}{*}{80\%} & POFRM & 10.05 (3.08) & \textbf{9.85 (3.35)} & \textbf{9.90 (3.07)} \\
 & POFRM-I & \textbf{9.867 (3.17)} & 10.37 (3.63) & 11.14 (3.68) \\
\hline
\end{tabular}
\caption{Mean and standard deviation (in parentheses) of 100 measures of the misclassification error for the scenarios where the response variable follows a Binomial distribution in the 1D simulation study.}
\label{Y_Binomial}
\end{table}

Tables \ref{Beta_1_Binomial} and \ref{Beta_2_Binomial} report the mean and standard deviation (in parentheses) of 100 IMSE measures for $\beta_1(t)$ and $\beta_{2}(t)$, respectively, under Binomial responses. The corresponding misclassification errors are shown in Table \ref{Y_Binomial}. Across all simulated conditions, the proposed POFRM method consistently performs on par with or better than the imputation-based variant (POFRM-I). This advantage becomes increasingly evident as the proportion of missing data or gap size grows—situations in which reconstruction-based approaches typically deteriorate.

For both functional coefficients, $\beta_1(t)$ and $\beta_2(t)$, POFRM attains lower mean IMSE values in most scenarios, with the gap between methods widening under higher missingness levels. In the few cases where POFRM-I yields slightly smaller errors, the differences are negligible, confirming that the direct estimation approach remains robust without requiring prior curve reconstruction.

Regarding predictive performance, both AUC values (Table \ref{AUC_Binomial_1D}) and misclassification errors (Table \ref{Y_Binomial}) support the same conclusion. Both methods exhibit high predictive accuracy overall, with AUCs close to 0.99 across all scenarios. However, POFRM generally preserves this level of discrimination even when data incompleteness reaches 60–80$\%$, while POFRM-I tends to show slightly higher variability and small degradations in classification accuracy, particularly at larger gap sizes.

Overall, the results demonstrate that directly modeling partially observed functional data through POFRM achieves predictive accuracy and functional parameter recovery comparable to—or better than—those obtained using imputation-based strategies, while avoiding the additional computational cost and potential bias introduced during the reconstruction step.

\begin{figure}[H]
    \centering
    \includegraphics[width=0.75\linewidth]{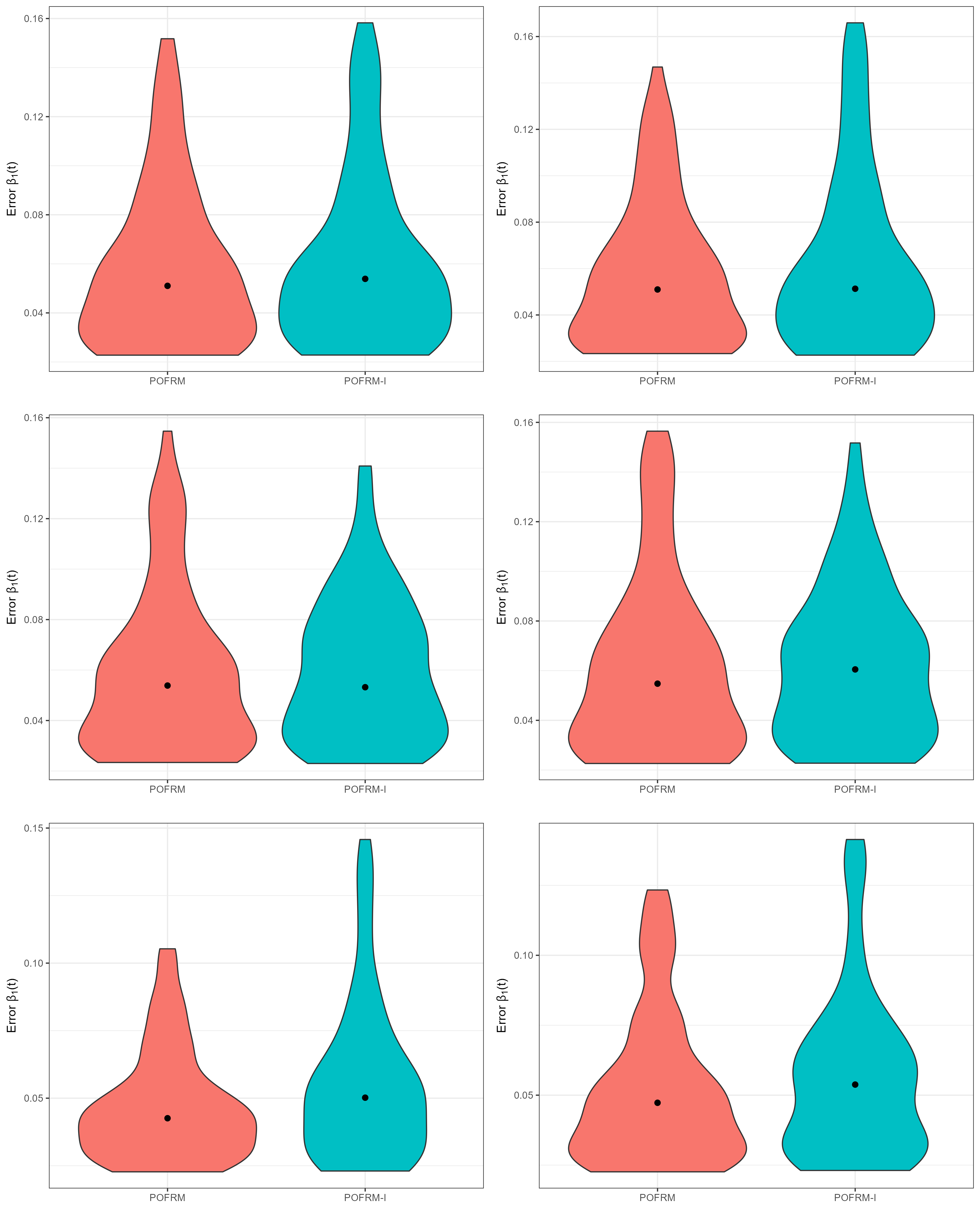}
    \caption{IMSE for the functional coefficient $\beta_1(t)$ for the 1D simulation study with Binomial response. The left column corresponds to scenarios with 60\% of curves containing gaps, and the right column to 80\%. Rows correspond (top to bottom) to gap sizes of 5\%, 10\%, and 15\%.}
    \label{Beta_1_Binomial_fig}
\end{figure}

\begin{figure}[H]
    \centering
    \includegraphics[width=0.75\linewidth]{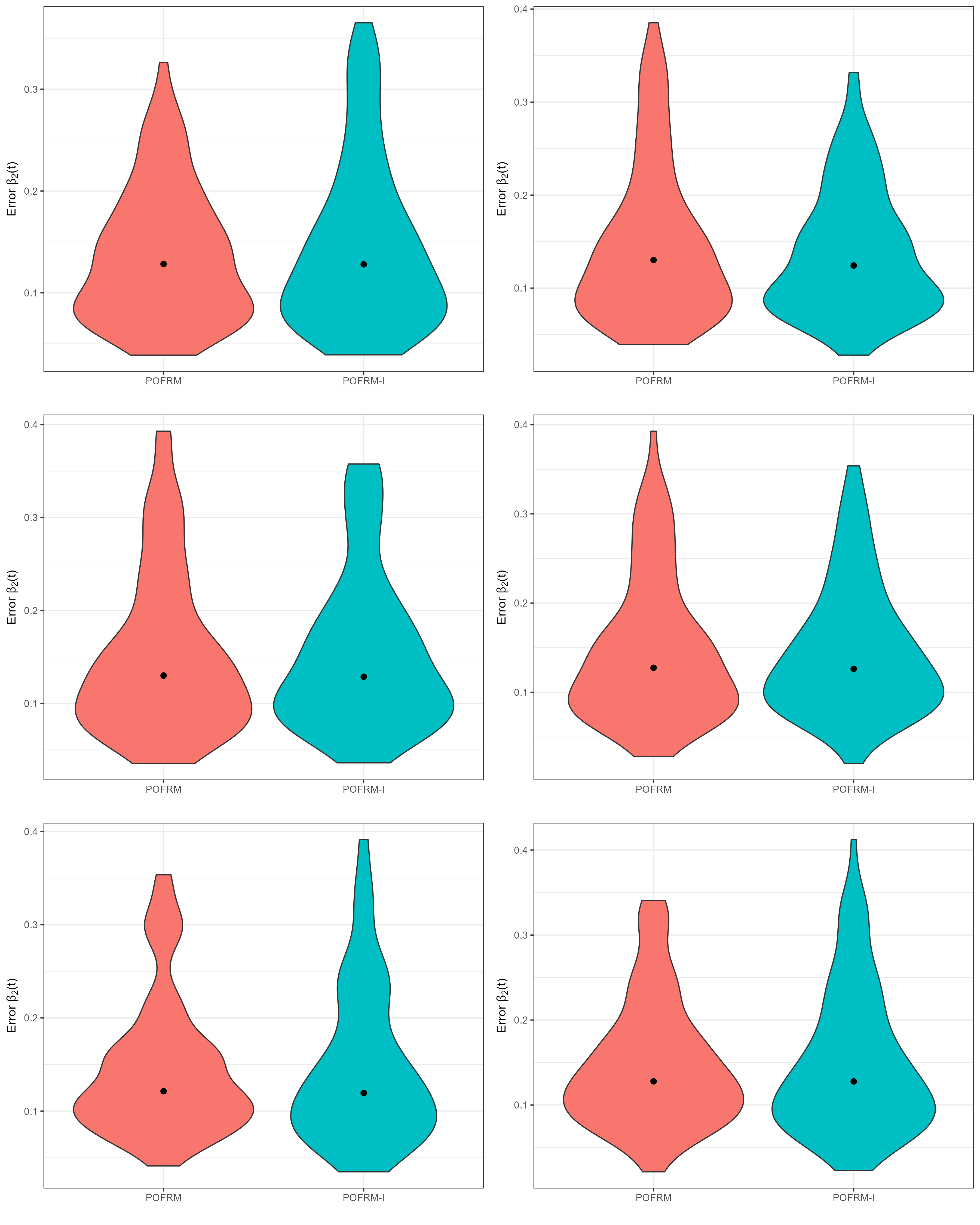}
    \caption{IMSE for the functional coefficient $\beta_2(t)$ for the 1D simulation study with Binomial response. The left column corresponds to scenarios with 60\% of curves containing gaps, and the right column to 80\%. Rows correspond (top to bottom) to gap sizes of 5\%, 10\%, and 15\%.}
    \label{Beta_2_Binomial_fig}
\end{figure}

\begin{figure}[H]
    \centering
\includegraphics[width=0.75\linewidth]{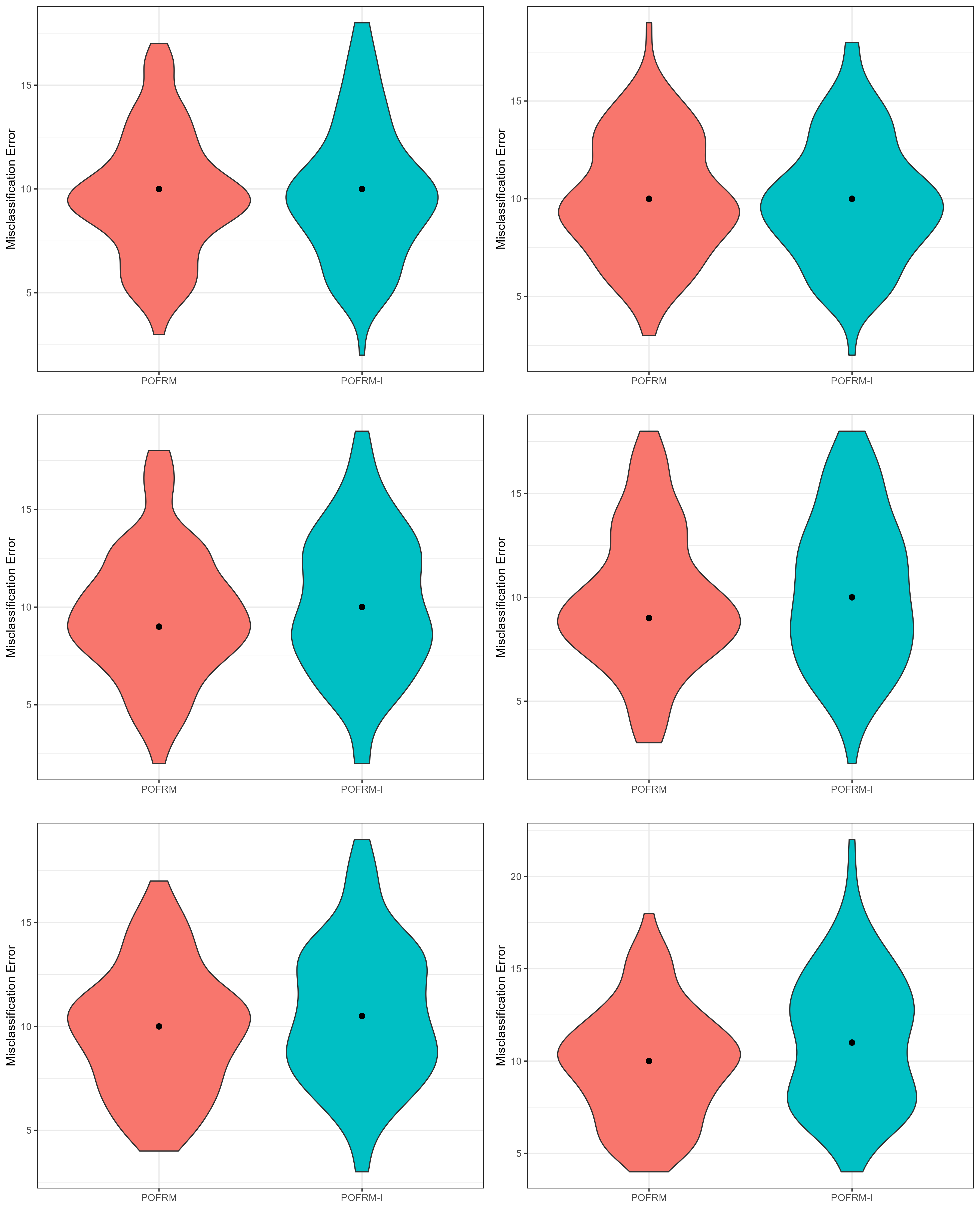}
    \caption{Misclassification error for the response variable for the 1D simulation study with Binomial response. The left column corresponds to scenarios with 60\% of curves containing gaps, and the right column to 80\%. Rows correspond (top to bottom) to gap sizes of 5\%, 10\%, and 15\%.}
    \label{Miss_fig}
\end{figure}

\subsection{Simulation study  2: two dimensional linear predictor}

This study focuses exclusively on models with functional covariates, extending the evaluation to the two-dimensional setting. A total of 18 scenarios were considered, each based on 100 simulated datasets.

\begin{itemize}
    \item \textbf{Outcome types}: As in the one-dimensional study, both continuous and binary outcomes were analyzed using the following predictor:

\begin{equation*}
    \eta_i = \alpha + \displaystyle \iint_{0}^{1} X_i(t_1,t_2) \beta(t_1,t_2)\text{d}t_1\text{d}t_2, \;\; t_1,t_2= [0,1]\times [0,1],
		\end{equation*}

 where the response variables were generated analogously to the 1D case: 
    \( Y_i = \eta_i + \epsilon_i \) with \( \epsilon_i \sim N(0, \delta_i) \) for the continuous outcome, and 
    \( Y_i \sim \text{Bernoulli}(p_i) \) with \( p_i = \frac{\exp(\eta_i)}{1 + \exp(\eta_i)} \) for the binary case. 
    The intercept was set to \( \alpha = 0.5 \).\\[3mm]

\item \textbf{Percentage of surfaces with missing observations:} 20\%, 40\%, and 60\%.\\[2mm]

    \item \textbf{Percentage of missing observations (gap size):} 5\%, 10\%, and 15\%.\\[2mm]

\item \textbf{Functional covariate simulation}: Functional covariates were generated under a two-group design. The sample was divided into two groups according to a target proportion, and each observation was assigned to one group. The true functional covariate for observation \(i\) was defined as:
\begin{equation*}
	X_{i}(t_1,t_2) = a_{1i} \cdot \sin(2\pi t_1) + a_{2i} \cdot \cos(2\pi t_2) + 0.5 \cdot (t_1 - 0.5) \cdot (t_2 - 0.5) + 2.0 + \delta_{i}(t_1,t_2)
\end{equation*}
where the coefficients $a_{1i}$ and $a_{2i}$ are drawn from group-specific distributions:
\begin{align*}
\text{Group 1:} \quad &a_{1i} \sim \mathcal{N}(1.5, 0.3), \quad a_{2i} \sim \mathcal{N}(1.2, 0.3)\\
\text{Group 2:} \quad &a_{1i} \sim \mathcal{N}(-1.2, 0.3), \quad a_{2i} \sim \mathcal{N}(-1.0, 0.3)
\end{align*}
 The domain \((t_1,t_2)\) consisted of a \(20\times20 = 400\)-point regular grid over \([0,1]\times[0,1]\), and the random noise term followed  
    \( \delta_i(t_1,t_2) \sim \mathcal{N}(0, \sigma_i^2) \), with \( \sigma_i^2 = 0.25\,\text{Var}(X_i(t_1,t_2)) \). 
    
    As in the 1D study, the noise variance was set to one-quarter of the functional covariate’s variance to ensure consistent variability across surfaces.\\[2mm]

 
\item \textbf{Functional coefficient}:

\end{itemize}

\begin{equation*}
    \beta(t_1,t_2) = \sin(2\pi t_1) \cdot \cos(2\pi t_2) + 0.8 \cdot (t_1 - 0.5) \cdot (t_2 - 0.5)
\end{equation*}

\subsection{Performance criteria}

In this second simulation study, particular attention is given to \textbf{computational efficiency}, as adapting the imputation method of \citet{Kraus2015ComponentsData} to a two-dimensional framework is expected to incur a substantial computational cost. Even within a modest \(20\times20\) grid, each surface must be vectorized into a 400-dimensional vector before processing, greatly increasing complexity.  

Due to these computational constraints, together with the previously discussed limitations on the proportion of missing data and gap sizes, the design of this simulation study focuses on moderate levels of incompleteness. A broader analysis without these restrictions is presented later in the real-data application section.

As in the 1D case, performance evaluation depends on the type of outcome variable:
\begin{itemize}
    \item For \textbf{binary outcomes}, predictive accuracy was assessed using the \textit{misclassification error} and the \textit{AUC}.  
    \item For \textbf{continuous outcomes}, predictive performance was evaluated using the \textit{mean RMSE}, defined as in the previous section.
\end{itemize}

In addition, for continuous responses, the model’s ability to recover the true functional coefficient \( \beta(t_1,t_2) \) was assessed using the \textit{Integrated Mean Square Error (IMSE)}:
\begin{equation*}
    IMSE = \iint_{0}^{1} \left[\beta(t_1,t_2) - \hat{\beta}(t_1,t_2)\right]^2\,\text{d}t_1\,\text{d}t_2,
\end{equation*}
where \( \hat{\beta}(t_1,t_2) \) denotes the estimated functional coefficient.

\subsubsection{Results}

This section presents the results of the two-dimensional (2D) simulation study. The same structure of tables, figures, and nomenclature as in the 1D case has been maintained for consistency.

Tables \ref{Y_table_Normal_2D} and \ref{Beta_table_2D} summarize the results for the Normal-response scenarios, reporting the RMSE of the response variable and the IMSE of the functional coefficient, respectively. For the Binomial-response scenarios, Tables \ref{AUC_table_2D} and \ref{Y_table_Binomial_2D} present the AUC values and misclassification errors. Computational times (in seconds) were also recorded for both approaches under all conditions and are summarized in Tables \ref{Time_table_Normal} and \ref{Time_table_Binomial}. Violin plots (included in the supplementary material) provide a visual comparison of the computation time distributions across repetitions.

\vspace{3mm}
\noindent\textbf{Normal response scenarios}

\vspace{2mm}

\begin{table}[htbp]
\centering
\small
\begin{tabular}{|l|l|c|c|c|}
\hline
Surfaces with gaps & Method & \multicolumn{3}{c|}{Gap size} \\
 & & 5\% & 10\% & 15\% \\ \hline
\multirow{2}{*}{20\%} & POFRM & 0.5927 (0.0836) & \textbf{0.6270} (0.065) & \textbf{0.6202} (0.0488) \\
 & POFRM-I & \textbf{0.5926} (0.0837) & 0.6277 (0.07) & 0.6208 (0.0485) \\
\hline
\multirow{2}{*}{40\%} & POFRM & 0.6183 (0.05) & \textbf{0.6232} (0.031) & \textbf{0.6194} (0.034) \\
 & POFRM-I & \textbf{0.6167} (0.049) & 0.6240 (0.032) & 0.6195 (0.035) \\
\hline
\multirow{2}{*}{60\%} & POFRM & \textbf{0.6376} (0.04) & 0.6311 (0.027) & 0.6205 (0.0262) \\
 & POFRM-I & 0.6377 (0.039) & \textbf{0.6310} (0.028) & \textbf{0.6203} (0.0263) \\
\hline
\end{tabular}
\caption{Mean and standard deviation (in parentheses) of 100 measures of the RMSE for the response variable $Y$ in the scenarios where the response variable follows a Normal distribution in the 2D simulation study.}
\label{Y_table_Normal_2D}
\end{table}

\begin{table}[htbp]
\centering
\small
\begin{tabular}{|l|l|c|c|c|}
\hline
Surfaces with gaps & Method & \multicolumn{3}{c|}{Gap size} \\
 & & 5\% & 10\% & 15\% \\ \hline
\multirow{2}{*}{20\%} & POFRM & \textbf{1.271} (0.0166) & 1.219 (0.0540) & \textbf{1.185} (0.0326) \\
 & POFRM-I & 1.272 (0.0166) & \textbf{1.218} (0.0548) & 1.186 (0.0324) \\
\hline
\multirow{2}{*}{40\%} & POFRM & \textbf{1.202} (0.049) & \textbf{1.175} (0.0188) & \textbf{1.18} (0.0246) \\
 & POFRM-I & 1.203 (0.049) & 1.1760 (0.0189) & 1.181 (0.0247) \\
\hline
\multirow{2}{*}{60\%} & POFRM & \textbf{1.175} (0.020) & \textbf{1.1740} (0.0184) & \textbf{1.182} (0.021) \\
 & POFRM-I & 1.176 (0.021) & 1.1741 (0.0189) & 1.1821 (0.0208) \\
\hline
\end{tabular}
\caption{Mean and standard deviation (in parentheses) of 100 measures of the IMSE for the functional coefficient $\beta(t_1,t_2)$ in the scenarios where the response variable follows a Normal distribution in the 2D simulation study.}
\label{Beta_table_2D}
\end{table}

\begin{table}[htbp]
\centering
\small
\begin{tabular}{|l|l|c|c|c|}
\hline
Surfaces with gaps & Method & \multicolumn{3}{c|}{Gap size} \\
 & & 5\% & 10\% & 15\% \\ \hline
\multirow{2}{*}{20\%} & POFRM & \textbf{10.07} (0.26) & \textbf{9.92} (0.18) & \textbf{10.93} (0.40) \\
 & POFRM-I & 26.54 (0.45) & 26.63 (0.31) & 27.73 (0.51) \\
\hline
\multirow{2}{*}{40\%} & POFRM & \textbf{10.26} (0.22) & \textbf{10.44} (0.18) & \textbf{10.79} (0.18) \\
 & POFRM-I & 28.10 (0.33) & 28.20 (0.25) & 28.25 (0.27) \\
\hline
\multirow{2}{*}{60\%} & POFRM & \textbf{10.62} (0.16) & \textbf{10.84} (0.20) & \textbf{11.24} (0.18) \\
 & POFRM-I & 29.75 (0.30) & 29.62 (0.32) & 29.73 (0.29) \\
\hline
\end{tabular}
\caption{Mean and standard deviation (in parentheses) of 100 measures of the computing time in seconds in the scenarios where the response variable follows a Normal distribution in the 2D simulation study.}
\label{Time_table_Normal}
\end{table}

The 2D scenarios with a Normal response confirm the consistent advantage of the proposed POFRM method. Tables \ref{Y_table_Normal_2D} and \ref{Beta_table_2D} report the RMSE and IMSE for the response variable and the functional coefficient, respectively, while Table \ref{Time_table_Normal} presents the mean and standard deviation of computational times.

Overall, both POFRM and POFRM-I perform similarly in terms of predictive and estimation accuracy, though POFRM attains slightly lower RMSE and IMSE values in most cases. Specifically, POFRM achieves lower RMSE in 5 out of 9 scenarios (56\%) and lower IMSE in 8 out of 9 scenarios (89\%), indicating greater stability in recovering the true functional coefficient. Although the differences in mean error are small, the standard deviations of POFRM are consistently similar or smaller, suggesting comparable precision at lower computational cost.

\begin{figure}[H]
    \centering
    \includegraphics[width=0.75\linewidth]{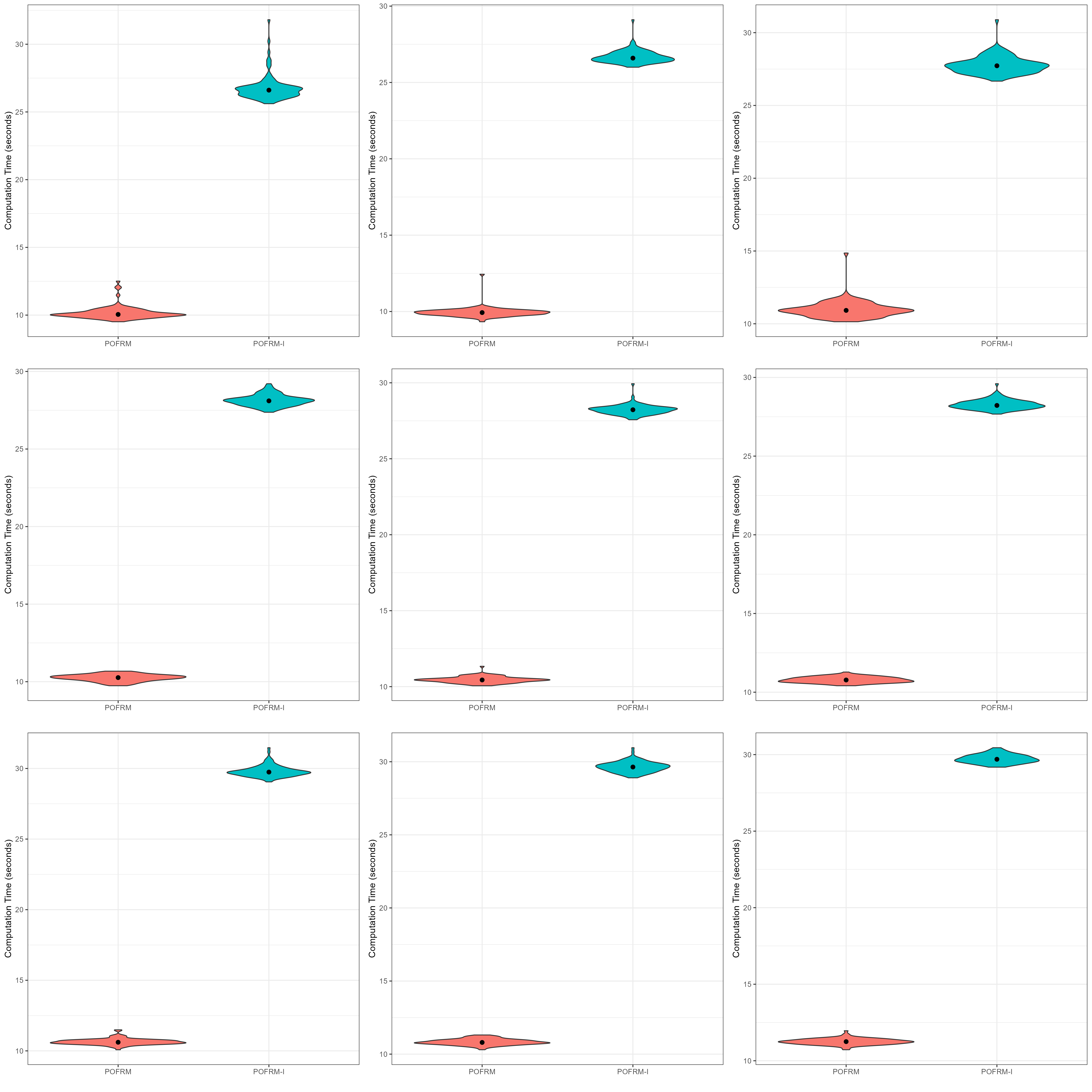}
    \caption{Computation time (in seconds) for the Normal-response scenarios in the 2D simulation study. Columns correspond (left to right) to gap sizes of 5\%, 10\%, and 15\%, while rows correspond (top to bottom) to proportions of surfaces with gaps of 20\%, 40\%, and 60\%.}
    \label{2D_Time_Normal}
\end{figure}

In contrast, the computational efficiency results show a pronounced difference between the two approaches. As Table \ref{Time_table_Normal} illustrates, POFRM is approximately three times faster than POFRM-I across all levels of missingness. While POFRM-I exhibits increased runtime as the proportion of missing surfaces or gap size grows, the POFRM method remains remarkably stable, evidencing its scalability and robustness. Figure \ref{2D_Time_Normal} visually reinforces this difference through the distribution of computing times across scenarios.

\vspace{3mm}
\noindent\textbf{Binomial response scenarios}

\vspace{2mm}

For the 2D simulation study with a Binomial response, the POFRM method again demonstrates overall superior performance compared to the filling-based strategy (POFRM-I). Tables \ref{Y_table_Binomial_2D} and \ref{AUC_table_2D} present the misclassification errors and AUC values, while Table \ref{Time_table_Binomial} details the computational times.

\begin{table}[htbp]
\centering
\small
\begin{tabular}{|l|l|c|c|c|}
\hline
Surfaces with gaps & Method & \multicolumn{3}{c|}{Gap size} \\
 & & 5\% & 10\% & 15\% \\ \hline
\multirow{2}{*}{20\%} & POFRM & \textbf{18.520} (3.702) & \textbf{18.750} (3.418) & 18.730 (3.360) \\
 & POFRM-I & 18.530 (3.691) & 18.790 (3.408) & \textbf{18.720} (3.361) \\
\hline
\multirow{2}{*}{40\%} & POFRM & \textbf{18.500} (3.577) & 18.660 (3.647) & \textbf{18.220} (3.424) \\
 & POFRM-I & 18.550 (3.600) & \textbf{18.570} (3.540) & 18.340 (3.508) \\
\hline
\multirow{2}{*}{60\%} & POFRM & \textbf{18.520} (3.445) & 18.880 (3.387) & 18.980 (3.573) \\
 & POFRM-I & 18.580 (3.453) & \textbf{18.870} (3.382) & \textbf{18.940} (3.564) \\
\hline
\end{tabular}
\caption{Mean and standard deviation (in parentheses) of 100 measures of the misclassification error for the scenarios where the response variable follows a Binomial distribution in the 2D simulation study.}
\label{Y_table_Binomial_2D}
\end{table}

\begin{table}[htbp]
\centering
\small
\begin{tabular}{|l|l|c|c|c|}
\hline
Surfaces with gaps & Method & \multicolumn{3}{c|}{Gap size} \\
 & & 5\% & 10\% & 15\% \\ \hline
\multirow{2}{*}{20\%} & POFRM & \textbf{0.8478} (0.0418) & \textbf{0.8479} (0.0370) & 0.8481 (0.0408) \\
 & POFRM-I & 0.8476 (0.0418) & 0.8475 (0.0378) & \textbf{0.8483} (0.0407) \\
\hline
\multirow{2}{*}{40\%} & POFRM & \textbf{0.8555} (0.0372) & \textbf{0.8499} (0.0348) & 0.8511 (0.0411) \\
 & POFRM-I & 0.8530 (0.0397) & 0.8498 (0.0349) & \textbf{0.8512} (0.0411) \\
\hline
\multirow{2}{*}{60\%} & POFRM & \textbf{0.8558} (0.0397) & \textbf{0.8489} (0.0363) & 0.8483 (0.0433) \\
 & POFRM-I & 0.8557 (0.0397) & 0.8488 (0.0364) & \textbf{0.8484} (0.0433) \\
\hline
\end{tabular}
\caption{Mean and standard deviation (in parentheses) of 100 measures of the AUC for the scenarios where the response variable follows a Binomial distribution in the 2D simulation study.}
\label{AUC_table_2D}
\end{table}

\begin{table}[htbp]
\centering
\small
\begin{tabular}{|l|l|c|c|c|}
\hline
Surfaces with gaps & Method & \multicolumn{3}{c|}{Gap size} \\
 & & 5\% & 10\% & 15\% \\ \hline
\multirow{2}{*}{20\%} & POFRM & \textbf{9.80} (0.32) & \textbf{10.89} (0.42) & \textbf{9.89} (0.17) \\
 & POFRM-I & 26.16 (0.39) & 27.85 (0.50) & 26.34 (0.28) \\
\hline
\multirow{2}{*}{40\%} & POFRM & \textbf{10.18} (0.23) & \textbf{11.05} (0.20) & \textbf{10.47} (0.18) \\
 & POFRM-I & 27.98 (0.38) & 29.06 (0.32) & 27.74 (0.23) \\
\hline
\multirow{2}{*}{60\%} & POFRM & \textbf{10.43} (0.20) & \textbf{11.45} (0.21) & \textbf{10.92} (0.17) \\
 & POFRM-I & 29.49 (0.31) & 30.53 (0.34) & 29.11 (0.26) \\
\hline
\end{tabular}
\caption{Mean and standard deviation (in parentheses) of 100 measures of the computing time (in seconds) in the scenarios where the response variable follows a Binomial distribution in the 2D simulation study. }
\label{Time_table_Binomial}
\end{table}

The differences between both methods in terms of predictive accuracy are modest but consistent. POFRM achieves lower misclassification errors in 5 out of 9 scenarios (56\%) and higher AUC values in 6 out of 9 scenarios (67\%), confirming slightly better predictive performance on average. The standard deviations are comparable, showing that both methods provide stable estimates.  

\begin{figure}[H]
    \centering
    \includegraphics[width=0.75\linewidth]{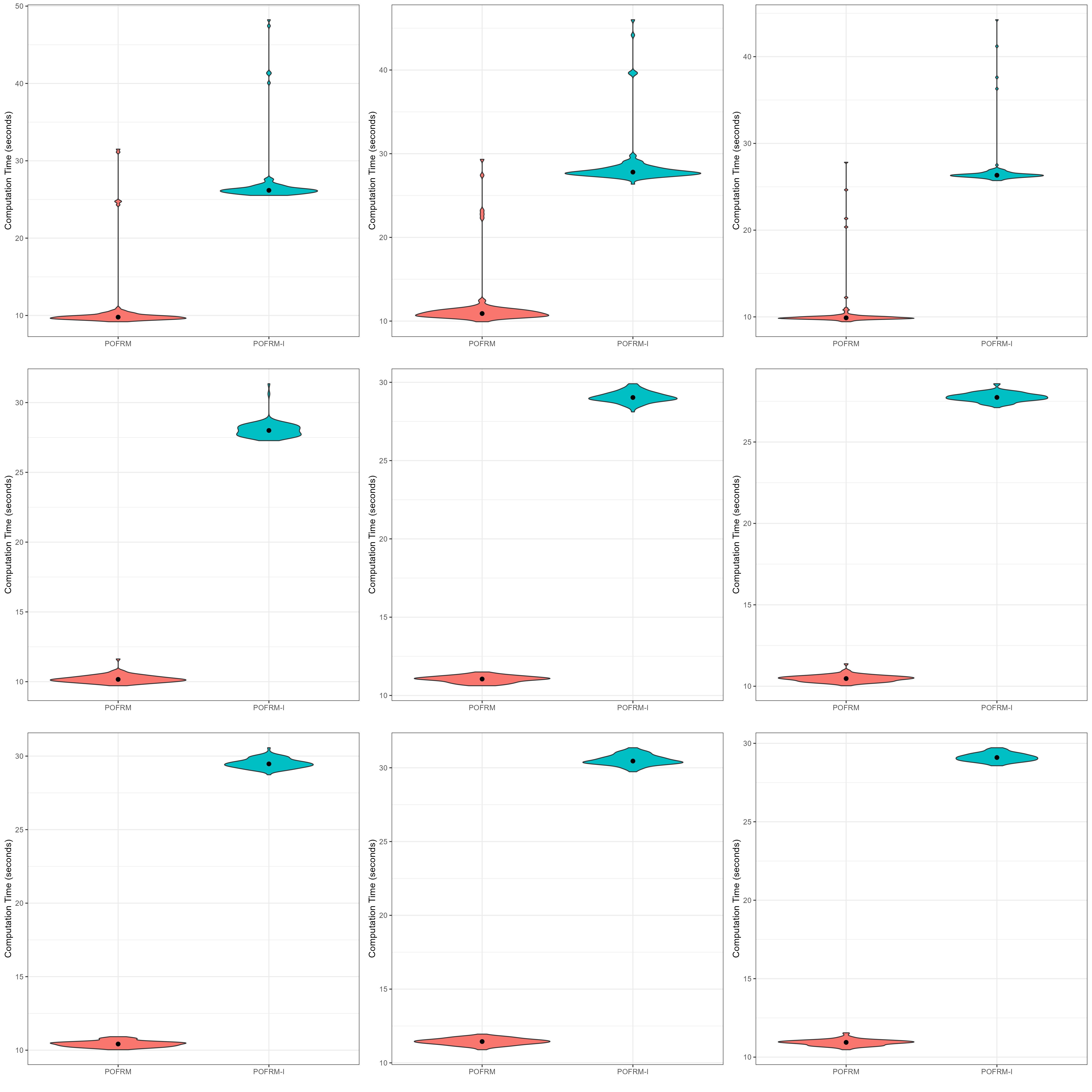}
    \caption{Computation time (in seconds) for the Binomial-response scenarios in the 2D simulation study. Columns correspond (left to right) to gap sizes of 5\%, 10\%, and 15\%, while rows correspond (top to bottom) to proportions of surfaces with gaps of 20\%, 40\%, and 60\%.}
    \label{2D_Time_Binomial}
\end{figure}

The most notable contrast appears again in computational efficiency. Across all scenarios, POFRM executes roughly three times faster than POFRM-I. The imputation-based approach requires an additional step to fill the missing regions, which becomes increasingly expensive as missingness grows. Since the POFRM-I method operates on vectorized surfaces of length 400 (corresponding to the $20\times20$ grids), this preprocessing step significantly inflates runtime. Conversely, the POFRM method directly accommodates partially observed data without reconstruction, resulting in far lower and more stable computational costs. Figure \ref{2D_Time_Binomial} displays the violin plots of computation times, highlighting this substantial efficiency gain.

Across both response types, the POFRM method proves to be the more efficient and robust alternative for modeling partially observed functional data in two-dimensional settings. Although both methods exhibit similar levels of predictive accuracy and functional recovery, POFRM attains these results at a fraction of the computational cost. The performance advantage becomes especially relevant as the proportion of missing data increases, where POFRM-I experiences a marked slowdown due to its imputation step.  

Overall, within the limitations imposed by the filling strategy (which constrain feasible gap sizes and percentages of incomplete surfaces), POFRM achieves better or comparable predictive results and substantially greater computational efficiency. Beyond these constrained scenarios, the comparison becomes impractical: the imputation-based approach either fails to run or requires prohibitive runtimes as missingness grows or the grid is refined. The following real-data application, conducted on a slightly larger domain, further illustrates this behavior and highlights the pronounced computational advantage of POFRM.

\section{Application to air quality classification}\label{section data}

To investigate the potential of functional regression models in environmental risk assessment, we analyze a dataset of sky images collected in Dimapur, India, for air quality classification. The study explores how functional representations of sky luminance patterns can serve as indicators of atmospheric pollution levels. The dataset includes 350 photographs labeled into two air quality categories—\emph{Moderate} (122 images) and \emph{Unhealthy for Sensitive Groups} (228 images)—and is publicly available at \href{https://www.kaggle.com/datasets/adarshrouniyar/air-pollution-image-dataset-from-india-and-nepal}{this
 link}.

\subsection{Data preprocessing}

Before applying the model, the images were preprocessed to obtain a functional representation suitable for analysis. Each image was converted from RGB color format to grayscale, standardizing the light intensity information while discarding color saturation, which may not directly relate to atmospheric conditions.  

Next, the images were cropped from their original $224 \times 224$ resolution to $70 \times 200$ pixels, focusing exclusively on the sky region. This step serves two purposes:  
\begin{enumerate}
    \item[i)] \textbf{Noise reduction:} removing non-relevant content (e.g., buildings, trees, ground) that could obscure the relationship between sky texture and air quality, and  
    \item[ii)] \textbf{Computational efficiency:} reducing dimensionality and hence the complexity of the functional model.  
\end{enumerate}

\begin{figure}[H]
    \centering
    \includegraphics[width=0.75\linewidth]{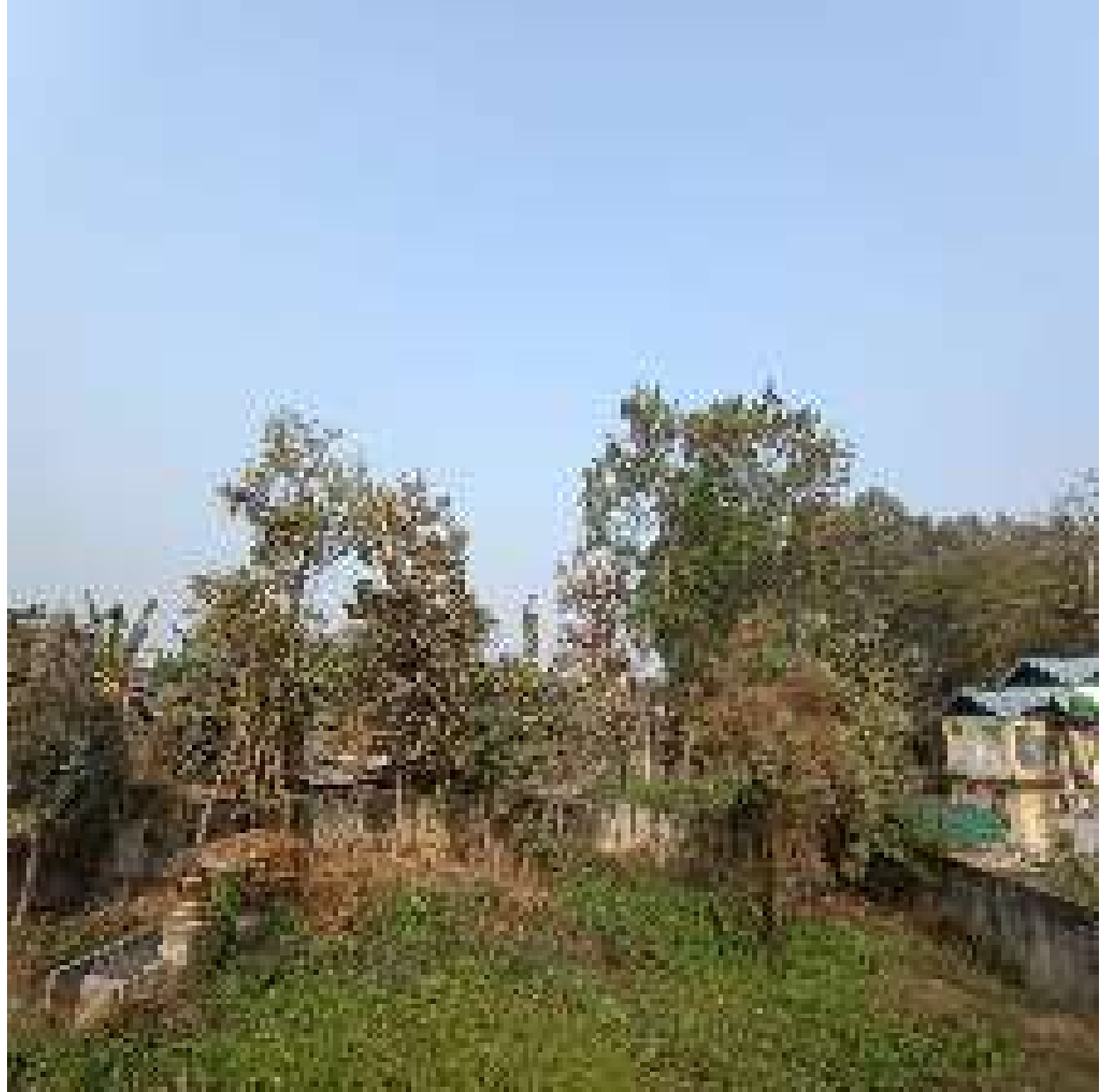}
    \caption{Example of a full image from the \emph{Unhealthy for Sensitive Groups} category.}
    \label{Big picture}
\end{figure}

\begin{figure}[H]
    \centering
    \includegraphics[width=0.5\linewidth]{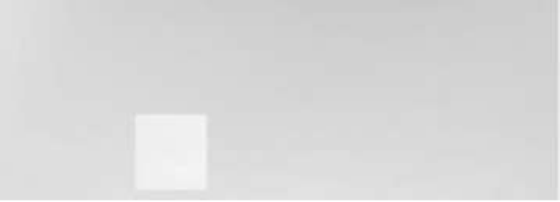}
    \caption{Preprocessed image showing only the sky region. The white area represents missing observations artificially introduced to simulate partial observation.}
    \label{Sky}
\end{figure}

Figure \ref{Big picture} shows an example of an original image for the \emph{Unhealthy for Sensitive Groups} category, and Figure \ref{Sky} illustrates the preprocessed version focusing on the sky portion.  
To test robustness under missingness, random gaps covering 10\% of each image were introduced to simulate partially observed data.

After preprocessing, the data can be expressed as $\{y_i, x_{ijk} = X_i(t_{1j_i}, t_{2k_i})\}$, for $i = 1, \ldots, 350$, where $t_{1j_i} \in D_i^1 \subset \{1,\ldots, 70\}$ and $t_{2k_i} \in D_i^2 \subset \{1,\ldots, 200\}$ represent the two-dimensional functional domain of each image.

\subsection{Results}

The confusion matrix in Table \ref{Image_table} summarizes the classification results. The proposed POFRM model achieved an accuracy of 0.9171, sensitivity of 0.9035, specificity of 0.9426, and an AUC of 0.97. These values indicate that the model can successfully discriminate between air quality categories based on subtle textural and brightness patterns in the sky, even with incomplete image data. In particular, the high AUC demonstrates strong discriminatory capability, while the balanced sensitivity and specificity suggest that the model performs consistently across both pollution levels.

\begin{table}[h!]
\centering
\small
\setlength{\tabcolsep}{10pt}
\renewcommand{\arraystretch}{1.3}
\begin{tabular}{|c|c|c|}
\hline
\multirow{2}{*}{\textbf{Predicted value}} & \multicolumn{2}{c|}{\textbf{True value}} \\ \hhline{|~|--|}
 & 0 & 1 \\ \hline
0 & \textcolor{red}{115} & 22 \\ \hline
1 & 7 & \textcolor{red}{206} \\ \hline
\end{tabular}
\caption{Confusion matrix for the air quality classification from POFRM model.}
\label{Image_table}
\end{table}

To allow direct comparison with the imputation-based POFRM-I method, we reduced the image size to $50 \times 50$ pixels, repeated the grayscale conversion, and randomly masked 15\% of each image. Additionally, only 20\% of the images were partially observed, replicating one of the simulation scenarios. The resulting confusion matrices for both methods are shown in Table \ref{Both_tables}, and a summary of performance metrics appears in Table \ref{comparison}.

\begin{table}[!h]
\centering
\small
\setlength{\tabcolsep}{10pt}
\renewcommand{\arraystretch}{1.3}
\begin{tabular}{|c|cc|c|cc|}
\hline
\textcolor{purple}{POFRM} method & \multicolumn{2}{c|}{True value} & \multicolumn{1}{l|}{\textcolor{teal}{POFRM-I} method} & \multicolumn{2}{c|}{True value} \\ \hline
Predicted value & \multicolumn{1}{c|}{0}                    & 1                    & Predicted value & \multicolumn{1}{c|}{0}                    & 1                    \\ \hline
0               & \multicolumn{1}{c|}{\textcolor{red}{117}} & 46                   & 0               & \multicolumn{1}{c|}{\textcolor{red}{114}} & 45                   \\ \hline
1               & \multicolumn{1}{c|}{5}                   & \textcolor{red}{182} & 1               & \multicolumn{1}{c|}{8}                    & \textcolor{red}{183} \\ \hline
\end{tabular}

\caption{Confusion matrices for air quality classification using reduced-resolution ($50 \times 50$) images, comparing the POFRM and POFRM-I methods.}
\label{Both_tables}
\end{table}

\begin{table}[!h]
\centering

\label{comparison}
\begin{tabular}{lcc}
\hline
\textbf{Metric} & \textbf{POFRM} & \textbf{POFRM-I} \\
\hline
Accuracy & 0.854 & 0.849 \\
Sensitivity & 0.959 & 0.934 \\
Specificity & 0.798 & 0.802 \\
AUC & 0.9119 & 0.9118 \\
Computation Time (s) & 591.63 & 21229.66 \\
\hline
\end{tabular}
\caption{Performance comparison between POFRM and POFRM-I methods on $50 \times 50$ pixel images.}
\end{table}

Both POFRM and POFRM-I exhibit comparable classification accuracy on reduced images; however, the POFRM method consistently performs slightly better in accuracy, sensitivity, and AUC, indicating more reliable identification of the ``Unhealthy'' class. This suggests that directly modeling partially observed surfaces without imputation allows the functional structure of the sky intensity field to be better preserved.  

From an environmental modeling perspective, the model effectively identifies patterns in sky luminance that are indicative of varying air pollution levels. By representing these spatial structures as smooth functional surfaces, the POFRM approach captures relevant spatial correlations that would otherwise be lost in pixel-based models. Furthermore, the most pronounced advantage of POFRM lies in its computational efficiency. Model estimation required approximately 9 minutes, compared to nearly 6 hours for the POFRM-I method. This dramatic difference arises because POFRM-I must first reconstruct missing data through imputation, vectorizing each surface (400-dimensional in the reduced setup) and solving large matrix systems. As image size or proportion of missing data increases, this step becomes computationally prohibitive. In contrast, POFRM integrates incomplete observations directly into the estimation process, avoiding the need for reconstruction and maintaining both speed and stability.

 These results indicate that the proposed methodology offers both statistical reliability and practical advantages for fitting functional regression models with partially observed data, particularly in two-dimensional contexts. In situations where the extent of missing information or the spatial resolution makes imputation-based strategies computationally demanding, POFRM provides an efficient and viable alternative. Even when the filling-based method can be applied, POFRM generally achieves comparable or better predictive performance while requiring substantially less computation time, highlighting its suitability for complex or data-intensive environmental applications.

\section{Conclusion}\label{section conclusion}

This work introduced a novel model for the analysis of partially observed functional data. The proposed approach is highly flexible, allowing for the inclusion of multiple functional covariates with potentially different dimensional structures. To the best of our knowledge, this is the first model capable of directly handling two-dimensional functional data without the need to impute missing values. In doing so, the methodology also provides a coherent and efficient strategy for smoothing functional covariates that are only partially observed, enabling accurate reconstruction and estimation without requiring ad hoc interpolation or imputation steps.  

The approach relies on a basis representation of both the functional covariates and the corresponding functional coefficients. The key idea is to evaluate the chosen basis over the full domain of the covariates and weight the resulting matrices with a structure that assigns zeros to positions corresponding to missing observations. This strategy allows for the estimation of the functional coefficients $a_i$ and ensures a proper numerical integration of the inner product matrices $\Psi_i^j$. As a result, the functional model is transformed into a multivariate one, which is then reparametrized within a mixed model framework to further enhance computational efficiency. By avoiding data completion, the proposed method achieves substantial reductions in computation time while maintaining high estimation accuracy.  

The performance of the model was thoroughly evaluated through two simulation studies involving one- and two-dimensional functional covariates. In both cases, our method was compared with a completion-based approach adapted from \cite{Kraus2015ComponentsData}. Since that technique was originally designed for one-dimensional data, we extended it to handle two-dimensional cases for a fair comparison. Across all scenarios, the proposed model consistently outperformed the competing method in terms of accuracy, robustness to missingness, and computational efficiency.  

To illustrate the method’s applicability, we applied it to the task of air quality classification in Dimapur, India, using partially observed sky-region images. The proposed model achieved high classification accuracy, low misclassification error, and a strong AUC value, demonstrating its capacity to extract meaningful information even from incomplete image data.  

Future research directions include extending the proposed framework to function-on-function regression models, particularly those involving missingness in either the predictors or the response functions.  

Finally, the proposed methodology has been implemented in the R package \textbf{VDPO}, now publicly available on the Comprehensive R Archive Network (CRAN). The most recent development versions, as well as the functions used for the simulation studies, can be accessed through the following \href{https://pavel-hernadez-amaro.github.io/VDPO/}{GitHub repository}.

\backmatter

\bmhead{Acknowledgements}

This work is supported by the grant PID2022-137243OB-I00. from the Spanish Ministry of Science, Innovation and Universities MCIN/AEI/10.13039/501100011033 and by CIAICO/2023/189 from the Conselleria de Educación, Universidades y Empleo de la Generalitat Valenciana.







\bibliography{sn-bibliography}

\end{document}